\documentclass[twocolumn]{aastex701}

\usepackage{mathtools}
\usepackage{graphicx} % Required for inserting images
\usepackage{xcolor}
\usepackage{soul}
\sethlcolor{white}
\newcommand{\bl}{}

\date{January 2026}

\shorttitle{Optimal Fluxes from Variable Illumination}
\shortauthors{Li et al.}

\received{\bl{January 15, 2026}}
\revised{\bl{March 20, 2026}}
\accepted{FUTURE DATE}
\submitjournal{Publications of the Astronomical Society of the Pacific}

\begin{document}

\title{Optimal and Unbiased Fluxes from Up-the-Ramp Detectors under Variable Illumination}

\correspondingauthor{Bowen Li}
\email{bowenj.li@mail.utoronto.ca}

\author[0009-0003-9321-0137]{Bowen Li}
\affiliation{Department of Statistical Sciences, University of Toronto, 700 University Avenue, Toronto, ON M5G 1Z5, Canada}
\email{bowenj.li@mail.utoronto.ca}

\author[0000-0001-7494-5910]{Kevin A. McKinnon}
\affiliation{David A. Dunlap Department of Astronomy \& Astrophysics, University of Toronto, 50 St George Street, Toronto, ON M5S 3H4, Canada}
\email{kevin.mckinnon@utoronto.ca} 

\author[0000-0002-6561-9002,gname=Andrew,sname=Saydjari]{Andrew~K.~Saydjari}
\altaffiliation{Hubble Fellow}
\affiliation{Department of Astrophysical Sciences, Princeton University,
Princeton, NJ 08544 USA}
\email{aksaydjari@gmail.com}  

\author[0000-0002-4454-1920]{Conor Sayres}
\affiliation{Department of Astronomy, University of Washington, Box 351580, Seattle, WA 98195, USA}
\email{csayres@uw.edu}

\author[0000-0003-3734-8177]{Gwendolyn M. Eadie}
\affiliation{Department of Statistical Sciences, University of Toronto, 700 University Avenue, Toronto, ON M5G 1Z5, Canada}
\affiliation{David A. Dunlap Department of Astronomy \& Astrophysics, University of Toronto, 50 St George Street, Toronto, ON M5S 3H4, Canada}
\email{gwen.eadie@utoronto.ca}  

\author[0000-0003-0174-0564]{Andrew R. Casey}
\affiliation{School of Physics and Astronomy, Monash University, Clayton, VIC 3800, Australia}
\affiliation{Center for Computational Astrophysics, Flatiron Institute, 162 Fifth Avenue, New York, NY 10010, USA}
\email{andrew.casey@monash.edu}

\author[0000-0002-9771-9622]{Jon A. Holtzman}
\affiliation{Department of Astronomy, New Mexico State University, P.O. Box 30001, MSC 4500, Las Cruces, NM 88033, USA}
\email{jholtzma@nmsu.edu}

\author[0000-0003-2630-8073]{Timothy D. Brandt}
\affiliation{Space Telescope Science Institute, 3700 San Martin Drive, Baltimore, MD 21218, USA}
\email{tbrandt@stsci.edu}

\author[0000-0003-3526-5052]{Jos\'e G. Fern\'andez-Trincado}
\affiliation{Universidad Cat\'olica del Norte, N\'ucleo UCN en Arqueolog\'ia Gal\'actica - Inst. de Astronom\'ia, Av. Angamos 0610, Antofagasta, Chile}
\affiliation{Universidad Cat\'olica del Norte, Departamento de Ingenier\'ia de Sistemas y Computaci\'on, Av. Angamos 0610, Antofagasta, Chile}
\email{jose.fernandez@ucn.cl}

\begin{abstract}
Near-infrared (NIR) detectors -- which use non-destructive readouts to measure time-series counts-per-pixel -- play a crucial role in modern astrophysics. Standard NIR flux extraction techniques were developed for space-based observations and assume that source fluxes are constant over an observation. However, ground-based telescopes often see short-timescale atmospheric variations that can dramatically change the number of photons arriving at a pixel. This work presents a new statistical model that shares information between neighboring spectral pixels to characterize time-variable observations and extract unbiased fluxes with optimal uncertainties. We generate realistic synthetic data using a variety of flux and amplitude-of-time-variability conditions to confirm that our model recovers unbiased and optimal estimates of both the true flux and the time-variable signal. We find that the time-variable model should be favored over a constant-flux model when the observed count rates change by more than 3.5\%. Ignoring time variability in the data can result in flux-dependent, unknown-sign biases that are as large as $\sim120\%$ of the flux uncertainty. Using real APOGEE spectra, we find empirical evidence for approximately wavelength-independent, time-dependent variations in count rates with amplitudes much greater than the 3.5\% threshold. Our model can robustly measure and remove the time-dependence in real data, improving the quality of data-model comparison. We show several examples where the observed time-dependence quantitatively agrees with independent measurements of observing conditions, such as variable cloud cover and seeing.
\end{abstract}


\section{Introduction} 
\label{sec:intro}

Extracting flux from near-infrared (NIR) observations is important for many areas of astrophysical research. NIR information, in particular, probes further into dusty regions than optical wavelengths, enabling the study of stars on the other side of our Galaxy or in the dust-obscured centers of neighboring galaxies. Many modern telescopes used in ongoing astronomical surveys use NIR detectors, such as the James Webb Space Telescope (JWST), the twin APOGEE spectrographs \citep{Wilson_2019} at the Apache Point \citep{Gunn_2006} and Las Campanas Observatories \citep{Bowen_1973}, the Prime Focus Spectrograph (PFS) on the Subaru telescope \citep{Sugai_2015}, and the soon-to-launch \bl{Nancy Grace Roman Space Telescope}. A key difference between NIR detectors and classical CCDs is that they can perform non-destructive readouts, which means that each pixel keeps a record of counts as a function of time. As a result, counts versus time plots tend to make a ``ramp'' shape, where the slope is defined by the incident flux. The time series associated with this measurement is often referred to as ``sampling up-the-ramp'' (SUTR).

A large amount of pioneering work on NIR detectors was done in preparation for JWST, with \citet{Fixsen_2000} first defining a model to estimate the flux and its corresponding uncertainty from SUTR observations. Subsequent work -- such as, \citet{Offenberg_2001} and \citet{Rauscher_2007} -- further refined the definitions of the expected fluxes and uncertainties and tested against real observations with NIR detectors. A key difference between these approaches comes from how the observed counts at different read times are weighted when measuring the best-fit slope/flux. \citet{Rauscher_2007}, for example, uses equal weights such that the best-fit slope is the one that minimizes the distance between all measured counts. This approach is statistically valid if noise is independent from read to read (as for idealized read noise), but does not account for the covariant behavior of photon noise.  \citet{Fixsen_2000}, however, applies larger weights to the first few and last few reads when photon noise dominates.  This leads to higher extracted signal-to-noise ratios (SNR), especially in the high flux limit. We note that the current JWST pipelines use the \citet{Fixsen_2000} approach while the main data reduction pipeline of the APOGEE spectrographs\footnote{\texttt{apogee\_drp} can be found here \url{https://github.com/sdss/apogee_drp/tree/daily}} \citep{Holtzman_2014,Nidever_2015} use the \citet{Rauscher_2007} approach. 

More recently, \citet{Brandt_2024} presented a new optimal fitting method that addresses many of the previous limitations. 
\cite{Fixsen_2000} used approximate, discrete, tabulated weights for combining the individual reads into fluxes. This sacrificed some SNR to avoid expensive computations associated with inverting a separate covariance matrix for every pixel. \citet{Brandt_2024}, by recasting the problem in the space of read differences, derived a closed-form solution for the maximum likelihood flux as a linear function of the individual reads. This enables unbiased and slightly higher SNR flux measurements at a small cost in computational performance. 
%In particular, this new approach works in the space of count differences between reads, and is able to measure unbiased and higher SNR fluxes compared to previous methods for a small cost in computation speed. 
%
%Because of some model assumptions, there are natural and expected correlations between the final flux measurements and their uncertainties, which manifests as small biases in Z-score (i.e. (measured flux $-$ truth)/uncertainty) at the faint flux end when using this method. 
Overall, the \citet{Brandt_2024} approach is very effective for space-telescope data where the flux of sources is constant over an exposure, such as for JWST and Roman. However, the constant-flux assumption begins to break down when dealing with ground-based telescopes, where atmospheric conditions can change substantially on the timescale of an exposure (i.e., a few minutes) or between different NIR detector readouts (i.e., $\sim10$~seconds). With variable illumination, using different weights from one pixel to another -- using either \citet{Fixsen_2000} or \citet{Brandt_2024} -- will lead to fluxes that are biased differently depending on the particular time-dependent signal and each pixel's true flux.  Any model that treats all pixels identically, like that of \citet{Rauscher_2007}, will avoid this differential bias but will sacrifice SNR. 

This paper proposes a new model to optimally extract fluxes from NIR detectors when dealing with time-variable data. Section~\ref{sec:statistics} presents the model and its underlying statistics. It explains the key underlying assumptions and caveats about its use. Section~\ref{sec:synthetic_testing} then tests the model with realistic synthetic data under a variety of conditions. These tests are expanded to real data from the northern APOGEE spectrograph in Section~\ref{sec:apogee_data}. Finally, our conclusions are presented in Section~\ref{sec:Conclusion} along with proposed future work focusing on improvements and implementation in a novel APOGEE data reduction pipeline. 

\section{Statistical Model} \label{sec:statistics}

Our goal is to measure, up to a multiplicative constant, unbiased relative fluxes of all pixels on a detector associated with a given spectrum read out up-the-ramp.  This could be accomplished by applying equal weights to the different reads in a pixel \citep[e.g.,][]{Rauscher_2007}, or by any model that applies the same read-by-read weights to all pixels. However, such an approach is not optimal.  Due to the different covariance of read noise and photon noise, higher flux SNRs are possible by weighting the reads differently depending on each pixel's level of illumination \citep{Fixsen_2000}. The best weights for each pixel depend both on the flux in that pixel and on the time-dependence of the illumination, an aspect that is not treated in either \cite{Fixsen_2000} or \cite{Brandt_2024}. We seek a generalization of the \citet{Brandt_2024} optimal approach that accounts for non-constant illumination, assuming the time dependence of the illumination is shared between neighboring pixels. In the case of \citet{Brandt_2024}, this illumination function is a constant. 

We begin by defining some terms. Let $n_r$ be the number of read-out times and $n_p$ be the number of pixels. For the case where the flux in a pixel is not changing over time, we expect that the observed counts in pixel $p$ at read $j$ is described by:
\begin{equation}
c_{j,p} = \delta_{j,p} + \sum_{i=1}^{j} \Delta r_{i,p}
\label{eq: synthetic data equation}
\end{equation}
where $j\in[1,n_r]$, $\delta_{j,p}$ is noise associated with the measurement, and $\Delta r_{i,p}$ is the true number of new photons that arrive at pixel $p$ between reads $j-1$ and $j$. \bl{In this way, we use $i$ to keep track of information \textit{between} reads while $j$ is used to track information \textit{at} different read times.} The measurement noise is described by:
\begin{equation}
    \delta_{j,p} \sim \mathcal{N}\left(\mu=0, \sigma^2=\sigma_{r,p}^2\right)
\end{equation} where $\sigma_{r,p}$ is the per-pixel read noise, which we assume has been previously measured during calibration of the detector \citep[e.g., using the model of][]{Brandt_2025}. The $\Delta r_{i,p}$ is described by:
\begin{equation}
    \Delta r_{i,p} \sim \mathrm{Poisson}\left(\lambda = f_p\right)
\end{equation} where $f_p$ is the true flux of the source in pixel $p$. 

Because modeling combinations of Poisson and Gaussian distributions is often computationally expensive, it is common to make a Poisson-approximated-as-Gaussian assumption such that:
\begin{equation}
    \Delta r_{i,p} \sim \mathrm{Poisson}\left(\lambda = f_p\right) \approx \mathcal{N}\left(\mu  =f_p, \sigma^2 = f_p\right),
\end{equation}
and we will do the same for this work. This assumption is very good in the case of large $f_p$ where Gaussian and Poisson distributions look quite similar. As $f_p$ decreases, the Poisson-to-Gaussian approximation is less good; however, this is often at the point where the impact on the observed counts from the read noise starts to dominate, so the final $c_{j,p}$ distribution is still well-approximated by a Gaussian for all $f_p$ values. After making this approximation, we are left with a combination of Gaussian distributions, which makes analysis much easier.  \citet{Brandt_2024} showed that if we change to work with count differences between neighboring reads instead of counts -- that is, $\Delta c_{j,p} = c_{j+1,p}-c_{j,p}$ -- then we have the following relationship:
\begin{equation} \label{eq:model_tb}
    \vec{\Delta c}_{p} \sim \mathcal{N}\left(\vec\mu = f_p \cdot \vec 1, \pmb V_p = \pmb V_{r,p} + f_p \cdot \pmb I\right)
\end{equation}
where $\vec{\Delta c}_{p}$ is a vector of the $(n_r-1)$ count differences, $\pmb I$ is the $(n_r-1)\times(n_r-1)$ identity matrix, and $\pmb V_{r,p}$ is a tri-diagonal matrix that described the read noise as
\begin{equation}
    \pmb{V}_{r,p} \;=\;
\scalebox{0.85}{$
\begin{bmatrix}
  2\sigma_{r,p}^2 & -\sigma_{r,p}^2 & 0 & \cdots & 0 \\
  -\sigma_{r,p}^2 & 2\sigma_{r,p}^2 & -\sigma_{r,p}^2 & \cdots & 0 \\
  0 & -\sigma_{r,p}^2 & 2\sigma_{r,p}^2 & \cdots & 0 \\
  \vdots & \vdots & \vdots & \ddots & \vdots \\
  0 & 0 & 0 & \cdots & 2\sigma_{r,p}^2
\end{bmatrix}.
$}
\end{equation}

Solving the model in Equation~\ref{eq:model_tb} is difficult in theory because the $f_p$ value of interest shows up in both the numerator and denominator of the arguments in the Gaussian likelihood. However, we often have a good first guess for what $f_p$ should be based on the data -- such as either the mean or median of the observed count differences -- especially in the case of a large number of reads, which is quite typical with NIR detectors. If we take this first guess, $\widehat f_p$, to generate $\widehat {\pmb V_{p}} = \pmb V_{p}(\widehat f_p)$, then the best fit flux is defined to be:
\begin{equation}
    \mu_{f_p} = \left(\vec{1}^T \cdot \widehat {\pmb V_{p}}^{-1} \cdot \vec{1} \right)^{-1} \cdot \left(\vec{1}^T \cdot\widehat {\pmb V_{p}}^{-1} \cdot \vec{\Delta c}_p \right)
\end{equation}
with uncertainty
\begin{equation}
    \sigma_{f_p} = \left(\vec{1}^T \cdot \widehat {\pmb V_{p}}^{-1} \cdot \vec{1} \right)^{-1/2}.
\end{equation}
We can then iterate using this new best guess for $f_p$ to update $\widehat f_p = \mu_{f_p}$ and $\widehat {\pmb V_{p}}$ to measure an improved $\mu_{f_p}$, repeating until convergence. In practice, \citet{Brandt_2024} find that a single update and recalculation is all that is needed to remove any bias in $\mu_{f_p}$. 

While this approach works successfully with space-based data where the fluxes of sources are truly quite stable over the course of an exposure, this is not often the case for ground-based observatories that have confounding effects like changing atmospheric conditions. To address this fact, we adjust the model in Equation~\ref{eq:model_tb} to include a time-varying component:
\begin{equation}
\begin{split}
    b_i &\geq 0\\
    f_p &\geq 0\\
    \Delta r_{i,p} &\sim \mathrm{Poisson}\left(\lambda = f_p \cdot b_i\right)\\
\end{split}
\label{eq:time_model}
\end{equation}
where $b_i$ is the ``flux multiplication factor'' for read difference $i$. We define $\vec b$ as a length $n_r-1$ vector of the multiplication factors that describe how the fluxes change as a function of time: 0 for ``no light received'' and 1 for ``maximum transparency/efficiency during the observation''. With this model, we assume that all pixels we analyze have been impacted by the same multiplicative factor at a given read time. \bl{We are also assuming that these multiplicative factors can be approximated by a constant between adjacent reads. Future work might explore using definitions that encode/enforce continuity in time, for example by defining a smooth time-dependent function, $b(t)$, that can be integrated between readout times to calculate $\vec b$.}

\begin{figure}
    \centering
    \includegraphics[width=\linewidth]{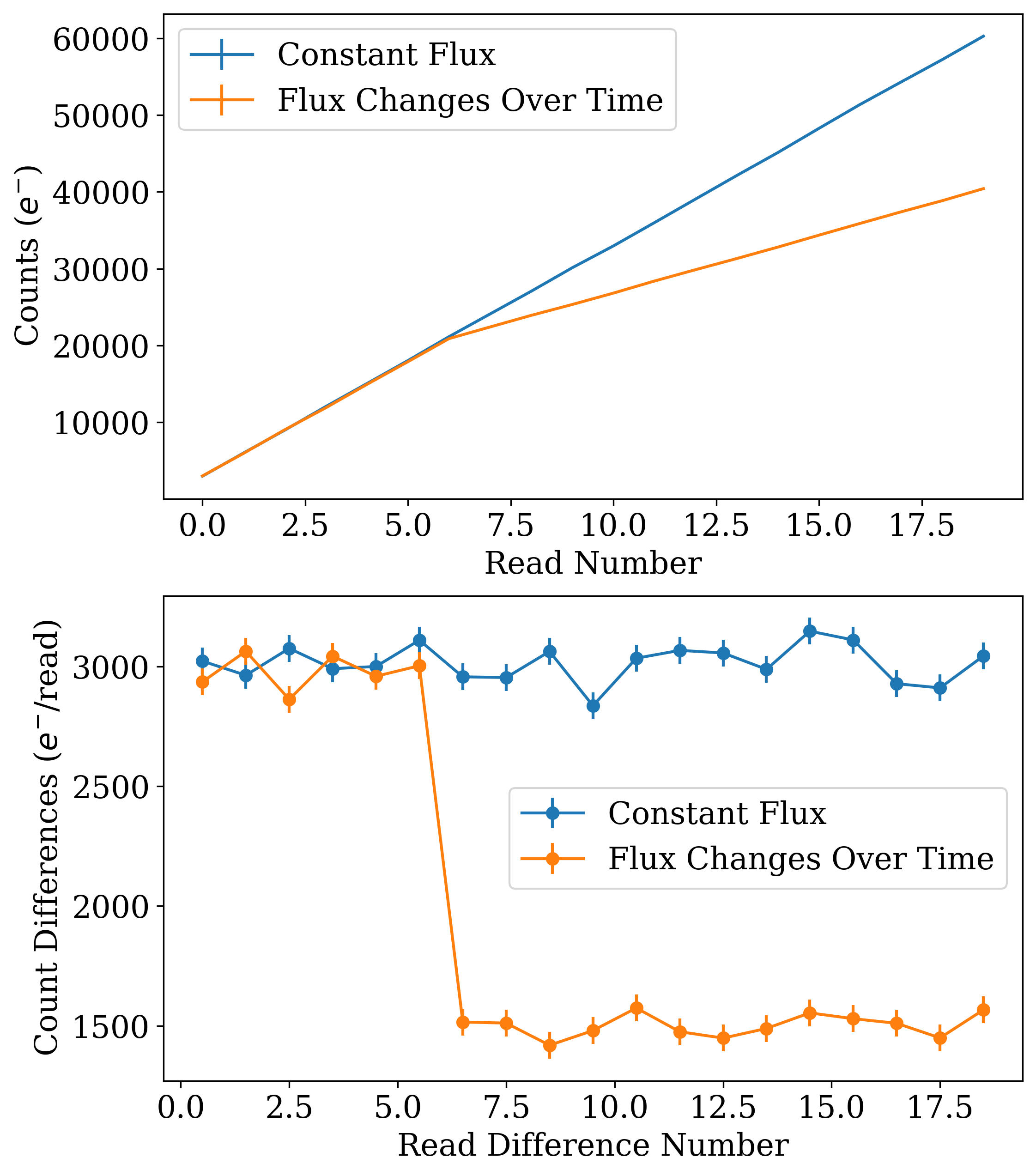}
    \caption{\textbf{Top:} An example ``ramp'' of counts versus read number using synthetic data. These data are generated using a Poisson distribution to simulate photon arrival with Gaussian read noise, as described in the beginning of Section~\ref{sec:statistics}. The blue line represents the observed counts in a pixel with constant flux over time, while the orange line shows the observed counts for flux that changes during the exposure to simulate the effects of, for example, a cloud moving in between reads six and seven.    \textbf{Bottom:} Count differences versus time using the same data as in the top panel. The true input flux for both cases is 3000 electrons per read, and the orange line's flux is decreased to half of its maximum after six reads.}
    \label{fig: Counts vs Read Times using synthetic data.}
\end{figure}

A comparison of example synthetic data generated using constant flux and a flux that is changing over time is shown in Figure~\ref{fig: Counts vs Read Times using synthetic data.}. For this synthetic pixel, the true flux is $f_p = 3000~e^-$\bl{/read} and the $\vec b$ for the orange line is 1 for the first six elements, and then 0.5 for subsequent read differences. One can imagine this represents a case where clouds rolled in for the later reads. It is apparent from both the counts and the count differences that a single flux will not describe the orange data. 

Using similar arguments to \citet{Brandt_2024}, we write the model as products of Gaussians such that:
\begin{equation} \label{eq:gaussian_time_model}
\vec{\Delta c}_{p} \sim \mathcal{N}  \left( f_p \cdot \vec{b}, \pmb{V}_p = \pmb{V}_{r,p} + \pmb{V}_{f,b,p} \right)
\end{equation}
where we now have
\begin{equation}
    \pmb{V}_{f,b,p} = f_p\cdot
\begin{bmatrix}
 b_1& 0 & \cdots & 0 \\
0 & b_2& \cdots & 0 \\
\vdots & \vdots & \ddots & \vdots \\
0 & 0 & \cdots & b_{n_r-1}
\end{bmatrix}.
\end{equation}
Because we assume that a collection of pixels share the same $\vec b$, we can jointly fit those pixels to constrain the time-varying component. For spectroscopic data, this usually means using pixels along the same trace of light, corresponding to a similar patch of sky (e.g., from one fiber or slit). For imaging data, this would mean using pixels that are near each other on the image. In this work, our assumption of a shared $\vec b$ between neighboring pixels means we are concerned with modeling wavelength-independent temporal variations. In Section~\ref{sec:apogee_data}, we will use real spectroscopic data to show that our achromatic model captures the vast majority of the time-dependent effects. Chromatic variations potentially contribute higher order effects that could be explored in future work.  

From the new model in Equation~\ref{eq:gaussian_time_model}, we notice that there is a degeneracy between $f_p$ and $\vec b$. In particular, we can arbitrarily multiply $f_p$ by any non-zero value so long as we divide $\vec b$ by the same factor. To fix this issue, we force one of the elements of $\vec b$ to be 1:
\begin{equation}
\begin{split}
    \vec{b}' = 
\begin{bmatrix}
1\\
b_2 \\
\vdots \\
b_{n_r-1}
\end{bmatrix} &= \begin{bmatrix}
1 \\
0 \\
\vdots \\
0
\end{bmatrix} + \begin{bmatrix}
0 & 0 & \cdots & 0 \\
0 & 1 & \cdots & 0 \\
\vdots & \vdots & \ddots & \vdots \\
0 & 0 & \cdots & 1
\end{bmatrix} \cdot \vec b\\
&= \vec{10} + \pmb{01} \cdot \vec b
\end{split}
\end{equation}
such that all other $\vec b$ elements are measured relative to this fixed one and $f_p$ is tied to that particular read time. In practice, we set the element in the first-guess estimate of $\vec b$ with the highest value to be 1 by rescaling $\vec b$ and $f_p$ appropriately. However, this choice does not require or enforce that all other $\vec b$ elements be smaller than one: in the fitting process, we might learn that one of the other elements is the true highest $\vec b$ element. Now, our full model becomes:
\begin{equation} \label{eq:linear_heirarchy}
    \begin{split}
        \vec{b} &\sim \mathcal{N}  \left( \vec b_{\mathrm{prior}}, \pmb V_{\mathrm{prior}} \right) \\
        \vec{\Delta c}_{p} &\sim \mathcal{N}  \left( f_p \cdot \vec b', \pmb{V}_p = \pmb{V}_{r,p} + \pmb{V}_{f,b',p} \right)\\
    \end{split}
\end{equation}
where $\vec{b}_{\mathrm{prior}} = \vec 1$
and
\begin{equation}
    \pmb {V}_{\mathrm{prior}}^{-1}  =
\lim _{a\to 0^+}\begin{bmatrix}
\frac{1}{a} & 0 & \cdots & 0 \\
0 & a & \cdots & 0 \\
\vdots & \vdots & \ddots & \vdots \\
0 & 0 & \cdots & a
\end{bmatrix}
\end{equation}
% \[
% \]
describe a non-informative prior that ensures our desired $\vec b$ element is held at 1 exactly: with this form, the fixed element of $\vec b$ has a prior that is infinitely narrow while the other elements have priors that are infinitely wide. After we define our flux vector as
\begin{equation}
    \vec{f} = 
\begin{bmatrix}
f_1\\
f_2 \\
\vdots \\
f_{n_p}
\end{bmatrix},
\end{equation}
we \bl{use the standard technique of taking a first partial derivative of the negative log likelihood, setting to 0 to find extrema, then evaluating the second derivative at that extremal value to determine the mean and covariance of the final posterior distribution. This results in} posterior conditionals for $\vec b$ and $\vec f$ described by:
% \begin{equation} \label{eq:posterior_conditionals}
% \begin{split}
%     (f_p \mid \vec{b}, \mathrm{data}) \sim \mathcal{N} \left(
% \begin{aligned}
% \sigma_{f,p}^2 &= \left[ (\vec{10} + \pmb{01}\cdot \vec{b})^T\cdot  \widehat{\pmb {V}}_p^{-1} \cdot (\vec{10} + \pmb{01}\cdot \vec{b})  \right]^{-1}, \\
% \mu_{f,p} &= \sigma_{f,p}^2 \left[ (\vec{10} + \pmb{01} \cdot\vec{b})^T \cdot \widehat{\pmb {V}}_p^{-1} \cdot \vec{\Delta c}_{p} \right]
% \end{aligned}
% \right)\\
% (\vec{b}\mid \vec{f} , \mathrm{data}) \sim \mathcal{N}\!\left(
% \begin{aligned}
% \pmb V_{b} &= \Bigl[ \sum_p f_p^2(\pmb{01}^T \cdot \widehat{\pmb {V}}_p^{-1} \cdot \pmb{01}) 
% + \pmb{V}_{\mathrm{prior}}^{-1} \Bigr]^{-1}, \\[6pt]
% \vec{\mu}_b &= \pmb V_b \cdot \Bigl[\sum_p 
%   \bigl(f_p \pmb{01}^T \cdot \widehat{\pmb {V}}_p^{-1} \cdot (\vec{\Delta c}_{p}- \vec{10}\cdot f_p)\bigr) \\[-2pt]
% &\qquad\quad + \pmb{V}_{\mathrm{prior}}^{-1} \cdot \vec{b}_{\mathrm{prior}} \Bigr]
% \end{aligned}
% \right)
% \end{split}
% \end{equation}
\begin{equation} \label{eq:posterior_conditionals}
\begin{split}
    (f_p \mid \vec{b}, \mathrm{data}) &\sim \mathcal{N} \left(\mu_{f,p}, \sigma_{f,p}^2\right) \\
    (\vec{b}\mid \vec{f} , \mathrm{data}) &\sim \mathcal{N}\left(\vec{\mu}_b ,\pmb V_{b}\right)
    \end{split}
\end{equation}
where
\begin{equation}
    \begin{split}
        \sigma_{f,p}^2 &= \left[ (\vec{10} + \pmb{01}\cdot \widehat{b})^T\cdot  \widehat{\pmb {V}}_p^{-1} \cdot (\vec{10} + \pmb{01}\cdot \widehat{b})  \right]^{-1}, \\
        \mu_{f,p} &= \sigma_{f,p}^2 \left[ (\vec{10} + \pmb{01} \cdot\widehat{b})^T \cdot \widehat{\pmb {V}}_p^{-1} \cdot \vec{\Delta c}_{p} \right],\\
        \pmb V_{b} &= \Bigl[ \sum_p \widehat f_p^2(\pmb{01}^T \cdot \widehat{\pmb {V}}_p^{-1} \cdot \pmb{01}) 
        + \pmb{V}_{\mathrm{prior}}^{-1} \Bigr]^{-1}, \\[6pt]
        \vec{\mu}_b &= \pmb V_b \cdot \Bigl[\sum_p 
          \bigl(\widehat f_p \pmb{01}^T \cdot \widehat{\pmb {V}}_p^{-1} \cdot (\vec{\Delta c}_{p}- \vec{10}\cdot \widehat f_p)\bigr) \\[-2pt]
        &\qquad\quad + \pmb{V}_{\mathrm{prior}}^{-1} \cdot \vec{b}_{\mathrm{prior}} \Bigr].
    \end{split}
\end{equation}
Here, we have used a similar trick as the \citet{Brandt_2024} method where we assume that we can estimate a good first guess for both $\widehat f$ and $\widehat b$ from the data to define $\widehat{\pmb V}_p$. By construction, $\vec \mu_b$ will be exactly 1 in its first element (or whichever has been chosen to be fixed) and will have a corresponding uncertainty of $\sigma_\mathrm{prior}$. Using these posterior conditionals, we can use a standard MCMC approach, such as Gibbs sampling, to draw samples from the posterior $(\vec f, \vec b \mid \mathrm{data})$ distribution. 

While Gibbs sampling of the distributions described in Equation~\ref{eq:posterior_conditionals} will give statistically robust samples of the desired posterior, it can be quite slow. If we instead recognize that we are almost always interested in simply measuring the maximum-a-posteriori (MAP) of the posterior, then getting the full shape of the distribution correct using Gibbs sampling is unnecessary. Instead, we can use a small number of iterations to get to the MAP using a linearized version of the model described by Equation~\ref{eq:gaussian_time_model}. In particular, we have the model count differences in pixel $p$ defined as 
\begin{equation}
    \vec m_p = f_p \cdot \vec b'= f_p (\vec{10} + \pmb{01}\cdot \vec b)
\end{equation} which has derivatives with respect to $\vec b$ and $f_p$ of
\begin{equation}
    \begin{split}
        \frac{\partial \vec m_p}{\partial \vec b} &= f_p \pmb{01}\\
        \frac{\partial \vec m_p}{\partial f_p} &= \vec{10}+\pmb{01}\cdot \vec b.\\ 
    \end{split}
\end{equation}
We can then define an offset vector from some initial guess $\vec v_{0}^T = \left(\vec { b_0}^T,\vec { f_0}^T\right)$ as
$$\vec{\Delta v} = \begin{bmatrix}
            \vec {\Delta b} \\
            \vec{\Delta f}
\end{bmatrix}$$ 
 that gives an approximation of the model count differences as:
\begin{equation}
    \begin{split}
        \vec m_p &\approx  f_{p,0} (\vec{10} + \pmb{01}\cdot \vec b_0) \\
        &\hspace{0.5cm}+\begin{bmatrix}
            \left(\frac{\partial \vec m_p}{\partial \vec b} \mid_{\vec v = \vec v_0}\right) & \left(\frac{\partial \vec m_p}{\partial \vec{f}} \mid_{\vec v = \vec v_0}\right)\\
        \end{bmatrix} \cdot \begin{bmatrix}
            \vec {\Delta b} \\
            \vec{\Delta f}
        \end{bmatrix}\\
        & := \vec m_{p,0} + \pmb U_p \cdot  \vec{\Delta v},\\
    \end{split}
\end{equation}
where $\pmb U_p$ is an $(n_r-1)\times (n_p + n_r-1)$ Jacobian matrix describing the first derivatives of $\vec b$ and $\vec f$. Here, $\frac{\delta \vec m_p}{\delta \vec{f}} \mid_{\vec f = \vec f_0}$ is an $(n_r-1)\times n_p$ matrix that is zeros everywhere except for column $p$, which is equal to the vector $\vec{10}+\pmb{01}\cdot \vec b_0$.
For clarity, an example set of $\pmb U_p$ matrices for $n_r = 4$, $n_p=2$, $\vec b^T = (1,b_2,b_3)$, $\vec f^T = (f_1,f_2)$ would look like:
\begin{equation*}
    \begin{split}
        \pmb U_1 = \begin{bmatrix}
        0 & 0 & 0 & 1 & 0 \\
        0 & f_1 & 0 & b_2 & 0 \\
        0 & 0 & f_1 & b_3 & 0 \\
        \end{bmatrix} \\
    \end{split}
\end{equation*}
and 
\begin{equation*}
    \begin{split}
        \pmb U_2 = \begin{bmatrix}
        0 & 0 & 0 & 0 & 1 \\
        0 & f_2 & 0 & 0 & b_2 \\
        0 & 0 & f_2 & 0 & b_3 \\
        \end{bmatrix}. \\
    \end{split}
\end{equation*}
Now, the linear-approximated statistical model becomes:
\begin{equation} \label{eq:linear_model}
    \begin{split}
        \vec{\Delta v} &\sim \mathcal{N}\left(\vec{\Delta v}_\mathrm{prior},\pmb V_{\vec{\Delta v},\mathrm{prior}}\right)\\
        \vec{\Delta c}_p &\sim \mathcal{N}\left(\vec m_{p,0} + \pmb U_p\cdot \vec{\Delta v}, \pmb V_p\right)
    \end{split}
\end{equation}
where again we have introduced a prior term that only serves to fix one of the $\vec b$ elements at exactly 1 \bl{with $\vec{\Delta v}_\mathrm{prior} = \vec 1$ (i.e., a vector of length $n_r-1+n_p$, where every element except the $\vec b$ entry we hold fixed could arbitrarily be set to any value) and }
\begin{equation}
    \pmb V_{\vec{\Delta v},\mathrm{prior}}^{-1}  =
\lim _{a\to 0^+}\begin{bmatrix}
\frac{1}{a} & 0 & \cdots & 0 \\
0 & a & \cdots & 0 \\
\vdots & \vdots & \ddots & \vdots \\
0 & 0 & \cdots & a
\end{bmatrix}.
\end{equation}
We note that the model in Equation~\ref{eq:linear_model} can be re-written to explicitly ignore the fixed $\vec b$ element and thereby remove the need for a prior -- in fact, this is how we implement the parameter fitting in our code -- but we leave it here for completeness. Combining the prior and likelihood above, we arrive at the following posterior distribution:
% \begin{equation}
%     \label{eq:linear_posterior}
%     \begin{bmatrix}
%             \vec {\Delta b} \\
%             \vec{\Delta f}
%         \end{bmatrix} \sim \mathcal{N} \!\left(
% \begin{aligned}
% \pmb V_{\vec {\Delta v}} &= \Bigl[ \sum_p (\pmb{U}_p^T \pmb {V}_p^{-1} \pmb{U}_p) 
% + \pmb{V_{prior}}^{-1} \Bigr]^{-1}, \\[6pt]
% \vec{\mu}_{\vec{\Delta v}} &= \pmb V_{\vec {\Delta v}} \cdot \Bigl[\sum_p 
%   \bigl(\pmb{U}_p^T \pmb {V}_p^{-1} (\vec{\Delta c}_{p}- \vec m_{p,0})\bigr) \\[-2pt]
% &\qquad\quad+ \pmb{V_{prior}}^{-1} \vec{\Delta v}_{prior} \Bigr]
% \end{aligned}
% \right).
% \end{equation}
\begin{equation}
    \label{eq:linear_posterior}
    \begin{split}
            \vec{\Delta v} \sim \mathcal{N} \left(\vec{\mu}_{\vec{\Delta v}}, \pmb V_{\vec {\Delta v}}\right)
    \end{split}
\end{equation}
with 
\begin{equation}
\pmb V_{\vec {\Delta v}} = \Bigl[ \sum_p (\widehat{\pmb{U}}_p^T \cdot \widehat{\pmb{V}}_p^{-1} \cdot \widehat{\pmb{U}}_p) 
+ \pmb{V}_{\vec{\Delta v},\mathrm{prior}}^{-1} \Bigr]^{-1},
\end{equation}
and
\begin{equation}
    \begin{split}
\vec{\mu}_{\vec{\Delta v}} &= \pmb V_{\vec {\Delta v}} \cdot \Bigl[\sum_p 
  \bigl(\widehat{\pmb{U}}_p^T \cdot \widehat{\pmb {V}}_p^{-1} \cdot (\vec{\Delta c}_{p}- \vec m_{p,0})\bigr) \\[-2pt]
&\qquad\quad+ \pmb{V}_{\vec{\Delta v},\mathrm{prior}}^{-1} \cdot  \vec{\Delta v}_{prior} \Bigr]    \end{split}
\end{equation}
In words, this new formalism says that we can use a starting guess ($\vec b_0$, $\vec f_0$), and then use gradient descent to calculate an improved subsequent guess, jointly fitting the $\vec b$ and $\vec f$. We then iterate this updating process until the parameters are no longer changing. It typically takes under five iterations to have all $\vec b$ elements changing by less than $1\times10^{-10}$ when the flux is relative high ($\geq100~e^-$\bl{/read}), with the $\vec f$ elements changing by a similar magnitude. Another approach for optimizing the fitting code could make use of, for example, \texttt{autodiff} to numerically estimate Jacobian vectors and Hessian matrices to quickly arrive at the MAP while also accounting for the $\pmb V_p$ dependence on $f_p$ and $\vec b$ in Equation~\ref{eq:linear_heirarchy}. 

A direct comparison between the results of Gibbs sampling and the linearized model are compared in Appendix~\ref{sec:method_comp_appendix}, with the key takeaway being that they agree very well in mean estimates of all parameters but show slight differences in the widths and correlations of their distributions. \bl{While the posterior conditional distributions of $\vec f$ and $\vec b$ are both multivariate Gaussians (i.e. Equation}~\ref{eq:posterior_conditionals}\bl{), the joint posterior is not exactly Gaussian (though it is close). This is because both $\vec f$ and $\vec b$ appear in the covariance matrices of each other's posterior conditional definitions. The small width and correlation differences in the joint $\vec b$ and $\vec f$ posterior distribution are likely a consequence of the linearized method approximating a close-but-not-entirely-Gaussian distribution as exactly Gaussian, with the Gibbs method giving more slightly more accurate estimates of uncertainties. However, we are almost always interested in the mean estimate of $\vec b$, which we can use to correct the data and then measure the resulting best-fit flux, as discussed in the following subsection. Because the mean estimates of $\vec b$ and $\vec f$ for the Gibbs and linearized methods are effectively equivalent, we are safe to choose either method.} For the remainder of the paper, all analyses will use the linearized model, as we have a fairly fast implementation that is similar to what most reduction pipelines would choose to employ. We also note that, should the Gibbs method be desired, the linearized or Hessian-based results provide a good parameter initialization because it will be at the mode of the posterior distribution. An implementation of our fitting procedure can be found at \url{https://github.com/KevinMcK95/fit_time_dep_NIR_fluxes}.

\subsection{Caveats and Recommendations} \label{sec:caveats}

Similar to the \citet{Brandt_2024} method, we have assumed we are able to estimate an initial guess for $\vec b$ and $\vec f$ from the data. In practice, this is much easier to do when the fluxes are substantially larger than the read noise and when using a fairly large number of pixels. We find that the linearized model is insensitive to the first guess parameters -- given enough iterations of the updates -- because the posterior distribution is unimodal. Our approach to define initializations for $\vec f$ and $\vec b$ is the following:
\begin{enumerate}
    \item Estimate $f_p$ by taking the median of the read differences in pixel $p$ (i.e. a good estimate if the flux was constant in time, $\vec b = \vec 1$);
    \item Divide the count differences by this $f_p$ to get an estimate of $\vec b$ for each pixel;
    \item Take an inverse-variance weighted average of those per-pixel $\vec b$ estimates to define the starting guess $\vec b$, where the weights come from the read noise and observed count differences in each pixel at each read time;
    \item Rescale the $\vec b$ guess such that the maximum of this vector is exactly 1;
    \item Divide the observed counts by this starting guess $\vec b$ and take an inverse-variance weighted mean of the corrected count differences in each pixel to define a starting guess $\vec f$.
\end{enumerate} 
To be consistent with physical expectations, we ensure that all the elements of $\vec f$ and $\vec b$ are positive at every step in the update process by replacing the updated best fit with 0 (or some very small non-zero threshold) if the current guess is negative. \bl{One could imagine working with log fluxes and log $\vec b$ elements or using more complicated prior distributions (e.g. truncated multivariate Gaussians) to ensure the posteriors always produce non-negative samples; however, these choices would break the posterior Gaussianity we rely on to perform computationally fast analyses. Future work could explore searching for relatively fast implementations using these non-negative constraints.}

When jointly fitting $\vec b$ and $\vec f$ -- either using the linearized model or Gibbs -- we end up with a distribution of $\vec f$ that is highly correlated between pixels. This correlation is often not desirable because later steps in processing a detector will typically make assumptions about pixel independence. Ideally, we could simultaneously fit all the pixels and keep track of all pixel correlations, but this would involve holding very large matrices in memory (i.e., greater than $n_\mathrm{pixels}\times n_\mathrm{pixels}$ in size). Similarly, tasks requiring the inversion of this large matrix would be impractical or impossible. For most applications, we recommend using the linearized model to jointly fit $\vec b$ and $\vec f$ using the available bright pixels. Then, throw away the joint $\vec f$ and keep only the best-fit $\vec b$. With this $\vec b_\mathrm{best}$, extract fluxes from all the desired pixels -- including those that were not used in the original joint fit -- using the posterior conditional distribution $(\vec f \mid \vec b = \vec b_\mathrm{best}, \mathrm{data})$ as defined in Equation~\ref{eq:posterior_conditionals}. In this way, the final fluxes will be uncorrelated with each other, preserving the desired pixel independence. For all the figures presented in this work, the fluxes we show are calculated using $(\vec f \mid \vec b = \vec b_\mathrm{best}, \mathrm{data})$ and not the fluxes we measuring during the joint fit. In this way, we can think of the $\vec f$ measured during joint fitting as nuisance parameters with $\vec b$ being the key measurement of interest. Of course, the ultimate goal is to measure the correct $(\vec f \mid \vec b = \vec b_\mathrm{best}, \mathrm{data})$. 

We note that the posterior $\vec b$ will also have correlations between the different elements corresponding to different read times. In theory, using $\vec b$ to accurately correct the observed data requires including the effect of these correlations by marginalizing over the posterior $\vec b$ distribution, $(\vec b \mid \mathrm{data})$ which we define to have covariance $\pmb V_{b,\mathrm{post}}$. However, this is quite costly and again introduces correlations between the pixel fluxes. Another approach is to approximate the marginalization by adding a $f_p^2 \pmb V_{b,\mathrm{post}}$ term to the data covariance matrix $\pmb V_p$ in $(\vec f \mid \vec b = \vec b_\mathrm{best}, \mathrm{data})$, and this is able to give back $\vec f$ uncertainties that are broadly consistent with results from the expensive marginalization. However, $\pmb V_{b,\mathrm{post}}$ is typically dense, so it breaks the fast-to-invert tri-diagonal form of $\pmb V_p$ and instead produces a slower-to-invert matrix that is potentially full-rank, significantly increasing computational costs. We explored using the full-rank matrix as well as only including the correlations up to the tri-diagonal elements when correcting the data: the full-rank version produces $\vec f$ uncertainties that are most consistent with the $\vec b$ marginalization results. To summarize, all cases of propagating $\vec b$ covariances only increased the magnitude of the $(\vec f \mid \vec b)$ uncertainties. As is shown in the Section~\ref{sec:synthetic_testing}, tests with synthetic data where we do not propagate $\vec b$ covariances -- that is, only the best fit $\vec b$ is used to measure $(\vec f \mid \vec b = \vec b_\mathrm{best}, \mathrm{data})$ -- return unbiased fluxes that agree with the input truth within their uncertainties.%, meaning that incorporating the $\vec b$ covariances leads to overestimated flux uncertainties. 
As a result, we recommend not propagating $\vec b$ uncertainties, assuming $\vec b$ can be measured to higher than $\sim10\%$ precision. 

In our model, we are assuming that the nonlinearity of each pixel has also been sufficiently accounted for through calibration of the detector. If non-linearity is still present in the data, then we expect that our model will pick up an average nonlinearity response from the jointly fit pixels: that is, even time-independent data will return a $\vec b$ with elements near 1 at the beginning read times and $< 1$ at later read times. However, we note that this is likely an undesirable way to handle nonlinearity corrections; there is no reason to believe that nearby pixels have the exact same nonlinearity response, but our model forces them to share a nonlinearity correction. It is much better to measure and remove nonlinearity before applying our model. 

When our model is applied to real data, we cannot know if $\vec b$ is solely measuring the effects of the sky. The $\vec b$ we measure could be a combination of improperly-removed nonlinearity effects, a changing sky, issues with telescope guiding, strange electronic noise, or persistence, to name a few. That being said, the source of the effect is not the greatest concern; it is clear that the data are poorly described by the \citet{Brandt_2024} constant flux model. As a result, such a model will have unaccounted-for systematic errors, underestimated uncertainties, and a flux-dependent bias. Our time variable model produces fluxes that are less biased, regardless of whether we understand the source of the effect. 

\bl{If external constraints of $\vec b$ were available -- for example, relatively strong prior information about the sky variability during an exposure from telescope guider cameras -- then disentangling the origin of different time variable signals would become more feasible. As we show later in Section}~\ref{sec:real_data_comp_with_guider}\bl{, guider camera information appears fairly correlated to the time variable signals we measure from real science data. Future work will focus on turning these guider camera estimates into well-calibrated prior constraints. After using telescope calibration data -- such as a combination of darks and dome flats -- to remove or model persistence and nonlinearity effects, the majority of the remaining time-variable signal could be described as a combination of sky and stellar variability. With well-informed priors on the sky variability, the $\vec b$ measurements could be decomposed to understand both the Earth's atmosphere as well as astrophysical variability in a particular star's brightness/spectrum. Thus, the methods presented in this paper may present a path towards learning interesting time-series stellar astrophysics over relatively short timescales ($\sim10$~seconds per read for APOGEE).}

In this work, we use the word ``bias'' to describe fluxes that have a non-zero offset in expectation from the truth. In the following sections, we show that fitting time-independent models to time-variable data produces flux-dependent biases, with the precise sign and magnitude depending on the particular $\vec b$ for an observation (Section~\ref{sec:fit_single_obs}, third panel of Figure~\ref{fig:synthetic_single_model_bias}). When we uniformly marginalize over many $\vec b$ draws (Section~\ref{sec:fit_repeat_obs_b_amp}), we find that the expectation of this bias goes to 0 (third panel of Figure~\ref{fig:synthetic_repeat_model_bias}), but this does not mean that the individual observations are unbiased. Instead, the overall effect of the bias can be thought of as a systematic error that is unaccounted for when using the time-independent model. Throughout this paper, the bias we mention is implicitly flux-dependent and ``systematic error'' and ``bias magnitude'' are used interchangeably. 

Measuring $\vec b$ is much easier when the count rate differences are significantly larger than read noise. However, many pixels -- whether in spectroscopic or imaging data -- will naturally be quite dark. For imaging data, we imagine a user would either measure a single common $\vec b$ for an image or calculate independent, local $\vec b$ values across the face of the detector using different bright sources, and then build some form of interpolation between those measurements to correct the data. For spectroscopic observations, the positions of bright sources/traces on the detector could be mapped to positions on the sky to allow the interpolation of $\vec b$ for faint sources. Another approach would be to focus on the flux near bright sky emission lines that should be present for all spectroscopic observations, regardless of the target's brightness. However, this latter approach has an additional complication in that the night sky is expected to change in different ways than sources over the course of the night; for example, the sky's large-scale trend of starting off bright, reaching a minimum, and then brightening towards the sunrise, or fast-timescale variations in hydroxyl radicals. Using skyline pixels to correct source pixels would then require calibrating the relative changing flux between a source and the sky, potentially harnessing prior constraints from independent measurements of the sky variability as a function of time (e.g., from guide/weather-tracking cameras sitting at the telescope). 

\begin{figure*}[t]
    \centering
    \includegraphics[trim={1cm 1cm 2cm 2cm}, clip, width=1.0\linewidth]{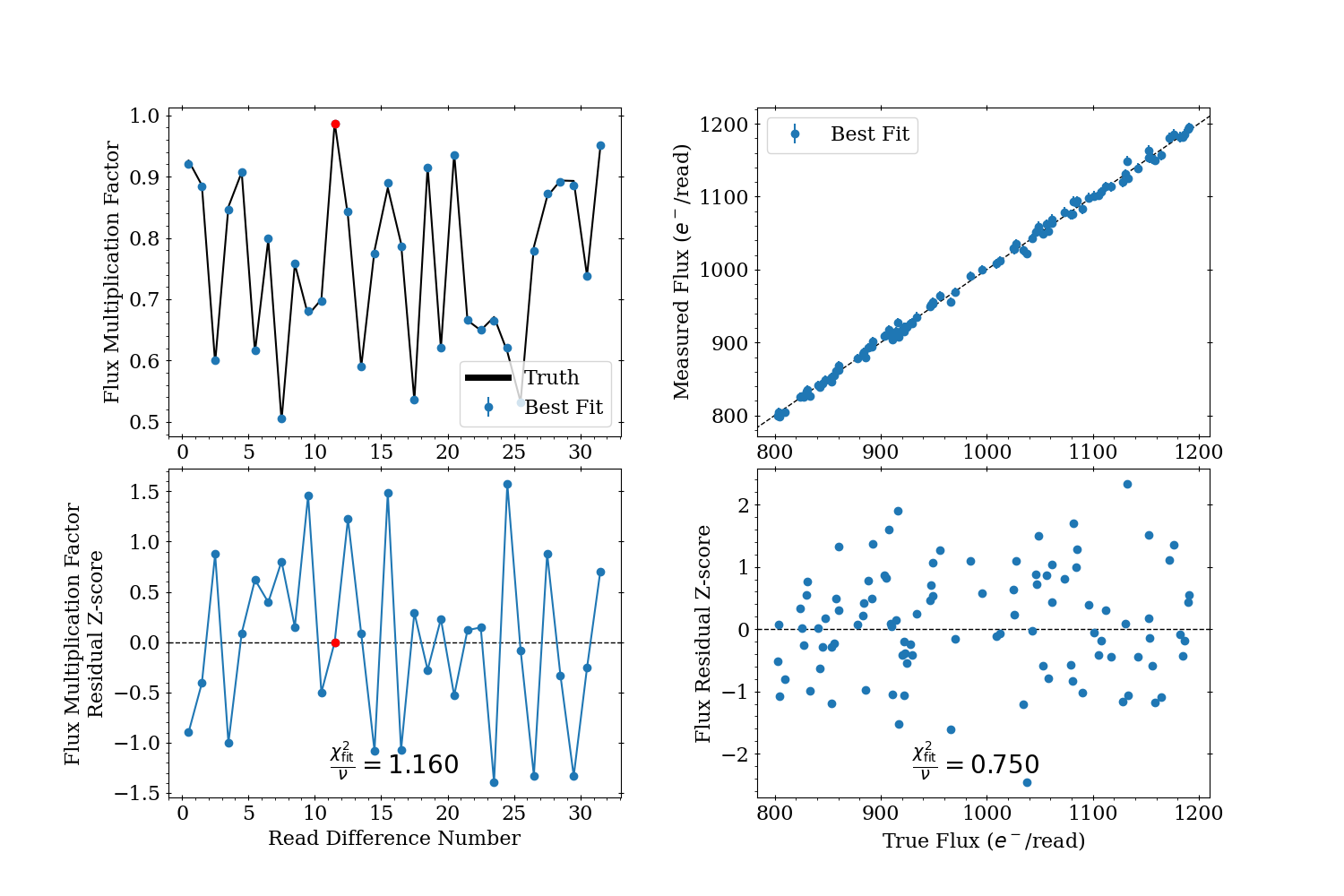}
    \caption{Comparison of the best-fit $\vec b$ and $(\vec f \mid \vec b= \vec b_{\mathrm{best}})$ and the truth for synthetic data from 100 pixels in a photon-dominated regime.  The left panels focus on $\vec b$, and the right panels focus on $(\vec f \mid \vec b)$. The top panels compare the input and outputs directly, while the bottom panels show the uncertainty-scaled residuals. The bottom panels display the best-fit $\chi^2$ scaled by the appropriate degrees of freedom, both of which are near 1.0 as expected for reasonable measurements and uncertainties. The red points in the $\vec b$ panels mark the read \bl{difference number} that was not used in the fitting to break degeneracies between $\vec b$ and $\vec f$, as discussed in Section~\ref{sec:statistics}.} \label{fig:synthetic_single_output_comp}
\end{figure*}

As mentioned earlier, there is an arbitrary tradeoff between $\vec b$ and $\vec f$, so it is important to remember that all the fluxes measured using the same $\vec b$ should be consistent with each other up to some multiplicative constant. However, when comparing pixels corrected with different $\vec b$ vectors, we have to be more careful. For spectroscopic data, we suggest measuring a single $\vec b$ per trace/fiber. Typically, the difference in relative fluxes between the traces is measured/calibrated at some later step in a data reduction pipeline, which should naturally handle differences between the arbitrary $\vec b$ scaling across the detector. How best to deal with the cross-talk regions between traces is not yet obvious, though the most statistically robust answer likely involves complete forward modeling of spectral light and point spread functions across the detector. For imaging data, locally measured $\vec b$ vectors should also be handled with care or a single $\vec b$ per image should be used. In both imaging and spectroscopy cases, it would be wise to force all independent $\vec b$ measurements to share the same reference read (i.e., where $b_i = 1$) so that everything is measured relative to the flux transparency at the same read time. 

Finally, we note that the analyses in the following sections regularly compare the results to the case where we assume the fluxes are constant in time, as measured by the \citet{Brandt_2024} model. While the results will show that our new model does a better job of describing our time-variable data, we must emphasize that we are not concluding that the \citet{Brandt_2024} model is incorrect. These comparisons merely show that ignoring time variability when it is present in the data leads to worse results. For data where the fluxes are truly stable over an observation -- such as for space telescopes -- the \citet{Brandt_2024} model is the correct choice to use. 

\section{Testing with Synthetic Data} \label{sec:synthetic_testing}

We generate synthetic data using the Poisson and Gaussian model defined in Equation~\ref{eq:time_model} given some choice of $\vec f$ and $\vec b$. In this section, we will focus on tests using a range of flux levels as well as $\vec b$ amplitudes to quantify our ability to measure unbiased fluxes.

\subsection{Fitting a Single Synthetic Observation} \label{sec:fit_single_obs}

We first test how well we can recover our input fluxes and $\vec b$ for a single realization of synthetic data. We set $n_{\mathrm{reads}} = 33$ and the read noise per pixel to be uniform random draws with bounds of $14-24~e^{-}$, which are typical values for APOGEE science observations. For this test case, we define the true $\vec b$ \bl{to} be $n_{\mathrm{reads}}-1$ uniform draws with bounds of $0.5-1.0$, the true fluxes \bl{to} be $n_{\mathrm{pixels}}$ uniform draws with bounds of $800-1200~e^-$\bl{/read}, and $n_{\mathrm{pixels}} = 100$. We have chosen 100 pixels to ensure that our fitting code (when used on a personal laptop) is able to run fast and is lightweight (in terms of RAM): future implementation with real data would \bl{likely} benefit from using all available pixels. Our current test represents a ``high'' flux case because the input fluxes are significantly larger than the read noise, making this the easiest test of our model. 

After generating the data, we estimate the first guess of the $\vec b$ and $\vec f$ using the approach described in Section~\ref{sec:caveats}, which is necessary to define the first covariance matrix that describe the data. We then use the starting guess $\widehat{\pmb V}_p$ matrices to update our best $\vec f$ and $\vec b$ guesses based on the linearized model of Equation~\ref{eq:linear_posterior}. We then update the $\widehat{\pmb V}_p$ matrices and repeat until all elements of $\vec b$ are changing less than some small tolerance (e.g., $1\times 10^{-10}$), which usually happens after only three or four iterations. 

\begin{figure}
    \centering
    \includegraphics[width=1.0\linewidth]{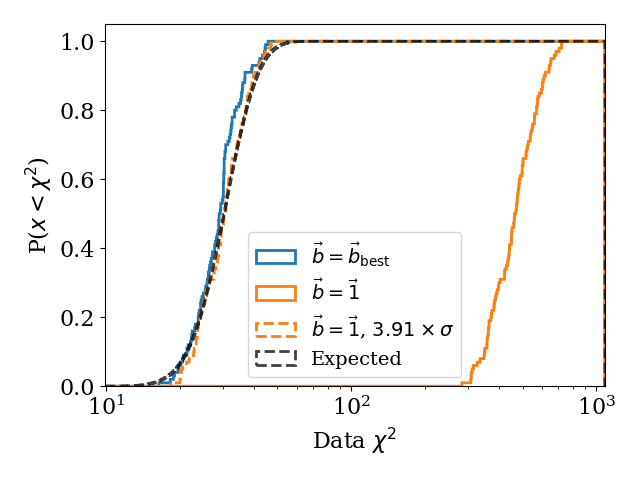}
    \caption{Comparison of the data $\chi^2$ empirical cumulative distribution function (CDF) for the same simulated data as in Figure~\ref{fig:synthetic_single_output_comp}. The blue line shows the $\chi^2$, as calculated using Equation~\ref{eq:chi2} when using the best fit $\vec f$ and $\vec b$, which shows good agreement with the expected $\chi^2_{\nu = n_\mathrm{reads}-2}$ distribution (black line). The solid orange line shows the data $\chi^2$ after measuring the best-fit constant flux following the \citet{Brandt_2024} model, which is not able to describe the data within its uncertainties. \bl{In fact, the solid orange histogram finds that the difference between the constant flux and data requires the uncertainties to be 3.91 times larger, as demonstated by the dashed orange histogram, which has much closer agreement to the expected curve.}}    \label{fig:synthetic_single_data_chi2}
\end{figure}

\begin{figure}
    \centering
    \includegraphics[width=1.0\linewidth]{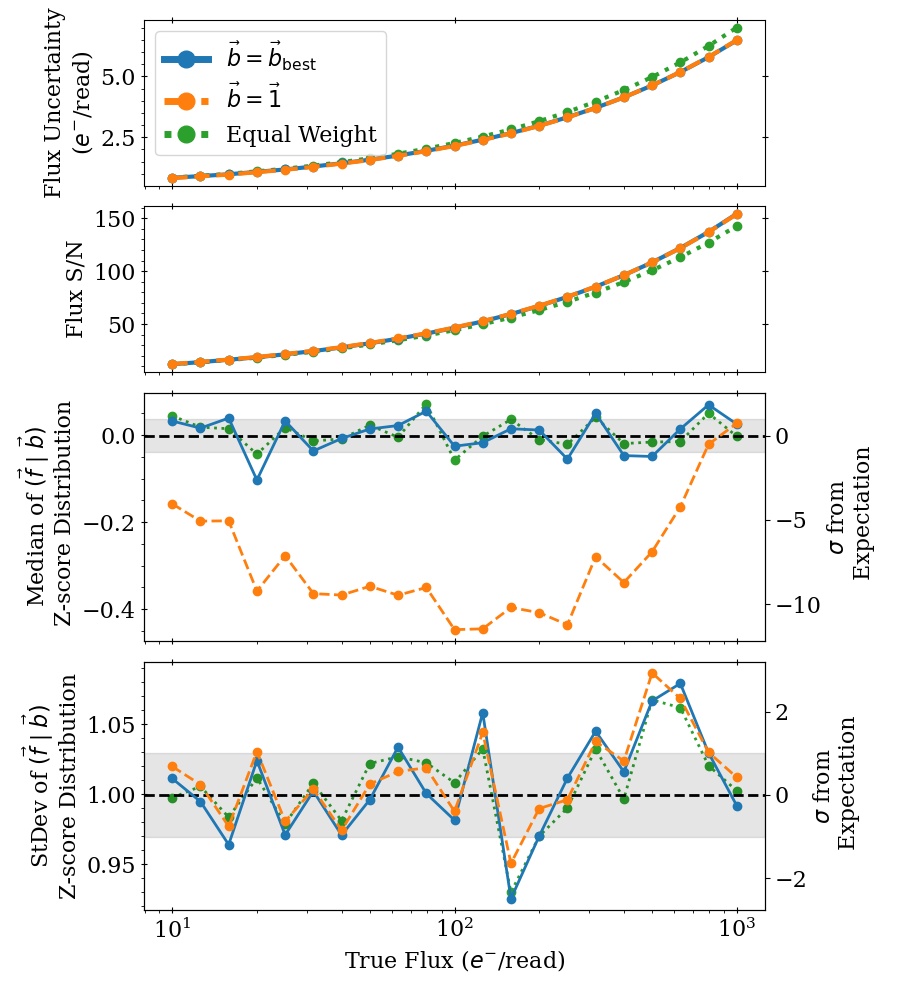}
    \caption{Comparison between the time-independent (orange lines) versus time-variable (blue lines) versus equal-weight (green lines) models for the extracted flux uncertainties (top), \bl{flux signal-to-noise ratios (second),} median biases (third), and residual distribution widths (bottom). Black lines and shaded regions correspond to the expected values and scatter given independent draws from a unit Gaussian. These measurements come from 1000 simulations of data using the same $\vec b$ as in Figure~\ref{fig:synthetic_single_output_comp}, but using a large range of flux levels. The top panel shows that time-variable and time-independent models predict extremely similar flux uncertainties (differences much smaller than 1\%) while the equal-weight model predicts larger uncertainties, especially at the high flux end. The \bl{third-from-top} panel emphasizes that neglecting to model time variability leads to a substantial flux-dependent bias when using the \citet{Brandt_2024} approach. The bottom panel shows that all the models have residual distributions with the expected widths.}    \label{fig:synthetic_single_model_bias}
\end{figure}

Before comparing the input and output $\vec f$ and $\vec b$, we must account for the arbitrary scaling between them. For example, if clouds in the sky meant that there was maximum $80\%$ of light getting through in a given exposure, then the true $\vec b$ is a vector of 0.8, but our model would return a vector of all ones for $\vec b$ with the $\vec f$ being a factor 0.8 lower than the true source fluxes. For real data, this discrepancy does not matter because we only ever expect to measure relative fluxes of sources, with later calibration steps determining an absolute fluxing if necessary. For our comparisons with synthetic data, however, we want to quantify how well our measurements and uncertainties describe the truth, which means we must rescale our outputs before we can compare with the input values. Specifically, we measure the multiplicative factor \bl{needed that} minimizes the difference between the true and measured $\vec b$, and then rescale the $\vec f$ and $\vec b$ by that amount before comparison.

The best-fit results for one realization of simulated data are shown in Figure~\ref{fig:synthetic_single_output_comp}, with the left panels focusing on $\vec b$ and the right panels focusing on $(\vec f \mid \vec b=\vec b_{\mathrm{best}})$. Both vectors have good agreement with the truth (upper panels) and any residuals are well-described by their uncertainties (bottom panels). We also calculate the data $\chi^2$ distribution using the best-fit fluxes and $\vec b$ using
\begin{equation}
    \label{eq:chi2}
    \chi^2 = \sum_p \left[ \vec{\Delta c}_{p} - f_p \cdot \vec b \right]^T \pmb {V}_p^{-1} \left[ \vec{\Delta c}_{p} - f_p \cdot \vec b\right],
\end{equation}
to confirm that the final best-fit model properly describes the data, and this is demonstrated in Figure~\ref{fig:synthetic_single_data_chi2}. We have also included the data $\chi^2$ distribution if we assume that the flux is constant in time and is extracted using the \citet{Brandt_2024} approach, which, as expected, is not able to describe the data within their uncertainties \bl{(solid orange line)}. In this case, the time-independent model requires the data uncertainties be inflated by a factor of 3.91 to recover the expected agreement \bl{(orange dashed histogram)}. 

We next test for bias between the fluxes measured using the time-independent and time-variable models. We do this by generating data 1000 times using the same true $\vec b$ as shown in Figure~\ref{fig:synthetic_single_output_comp}, but now across a range of flux levels that are typical for pixels in an APOGEE exposure. For each data draw, we extract fluxes using both $(\vec f \mid \vec b = \vec 1)$ and $(\vec f \mid \vec b = \vec b_\mathrm{best})$, where $\vec b_\mathrm{best}$ is the same best-fit value we have measured previously; that is, we do not re-estimate $\vec b_\mathrm{best}$ using the new data draws. We also extract fluxes using the equal-weight-per-read approach of \citet{Rauscher_2007}, where the fluxes come from their Equation 3 and the uncertainties are defined by their Equation 1. To be fair in our comparisons, we allow all the extracted fluxes in a realization to be scaled by a single multiplicative factor to have the best agreement between the model and the truth. At each flux level, we end up with a residual distribution whose median should be near zero and $\frac{1}{2}\cdot 68\%$ width should be near 1 to agree with the unit Gaussian. The result of this comparison is shown in Figure~\ref{fig:synthetic_single_model_bias}, with the top panel showing the extracted flux uncertainties, \bl{the second panel showing the flux signal-to-noise ratios,} the \bl{third} panel showing the median of the uncertainty-scaled residuals, and the bottom panel showing the width of the residual distribution. The top panel shows that both the time-variable and time-independent models (blue and orange lines) predict similar flux uncertainties as a function of true flux, with at maximum $\sim1\%$ differences between the two lines. The equal-weight model, however, predicts larger flux uncertainties, especially in the high flux limit ($\sim 8\%$ smaller SNR at a flux of $1000~e^-$\bl{/read}). The bottom panel shows that all models measure widths of their residual distributions that are well-described by their uncertainties; all the width lines also agree reasonably well with the black line and shaded region showing the expected mean and scatter based on the same number of draws from a unit Gaussian. The blue and green lines in the \bl{third} panel shows that both the time-variable and equal-weight models are able to recover the expected agreement with a zero median. However, the orange line shows that neglecting to account for time variability in the data leads to a systematic source of scatter in the constant flux model that is also nefariously flux-dependent: in this case, the systematic error is as large as $\sim0.45$ of the flux uncertainty in this case. When we repeat these tests with different realizations of $\vec b$ -- as we will explore in Section~\ref{sec:fit_repeat_obs_b_amp} -- we also find that the strength and sign of this bias can change significantly. This is a particularly pernicious type of bias because there is no simple way to correct for it after the fact. The only way to remove this bias and measure optimal flux uncertainties is to use the time-dependent model.

\begin{figure}[t]
   \centering
    \includegraphics[width=\linewidth]{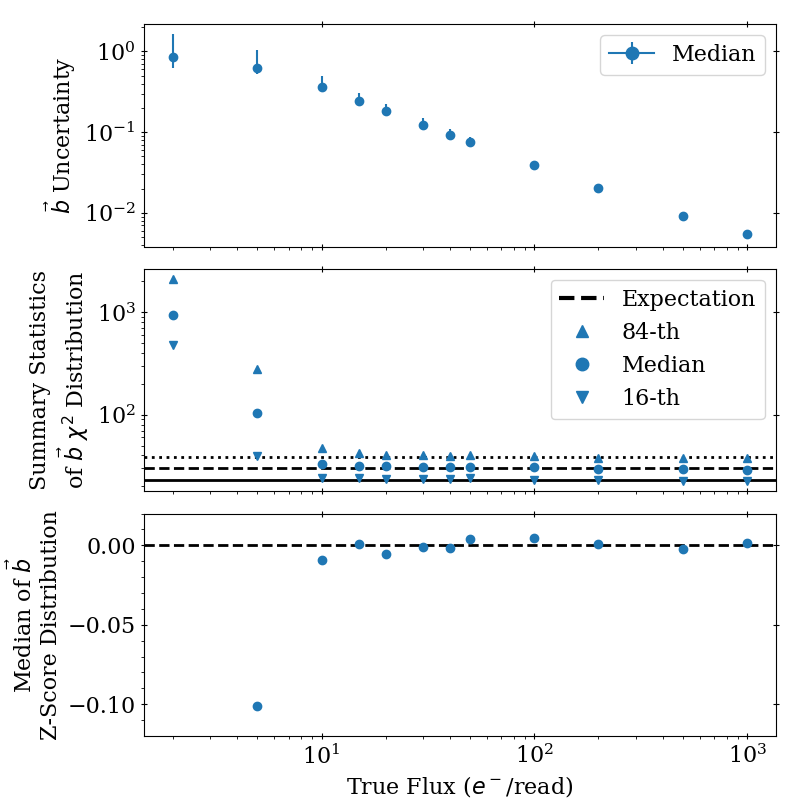}
    \caption{Summary of $\vec b$ measurements from repeated simulation across a range of flux levels using $n_\mathrm{reads} = 33$ and $n_\mathrm{pixels} = 100$. Each data point comes from 1000 iterations of fitting data with the same true $\vec b$ as shown in Figure~\ref{fig:synthetic_single_output_comp}. The top panel shows the median uncertainty in the measured $\vec b$ from all realizations\bl{, with the errorbars showing the 68\% interval from those realizations}. The middle panel shows the 16-th, 50-th, and 84-th percentiles of the $\chi^2$ distribution of measured $\vec b$ compared to the truth. The bottom panel shows the median of the $\vec b$ uncertainty-scaled residual distribution. This figure reveals that flux levels $>10~e^-$\bl{/read} are able to measure unbiased $\vec b$ using 100 pixels.}
    \label{fig:Chi_distribution_of_b_vector}
\end{figure}

\begin{figure}
    \centering
    \includegraphics[width=\linewidth]{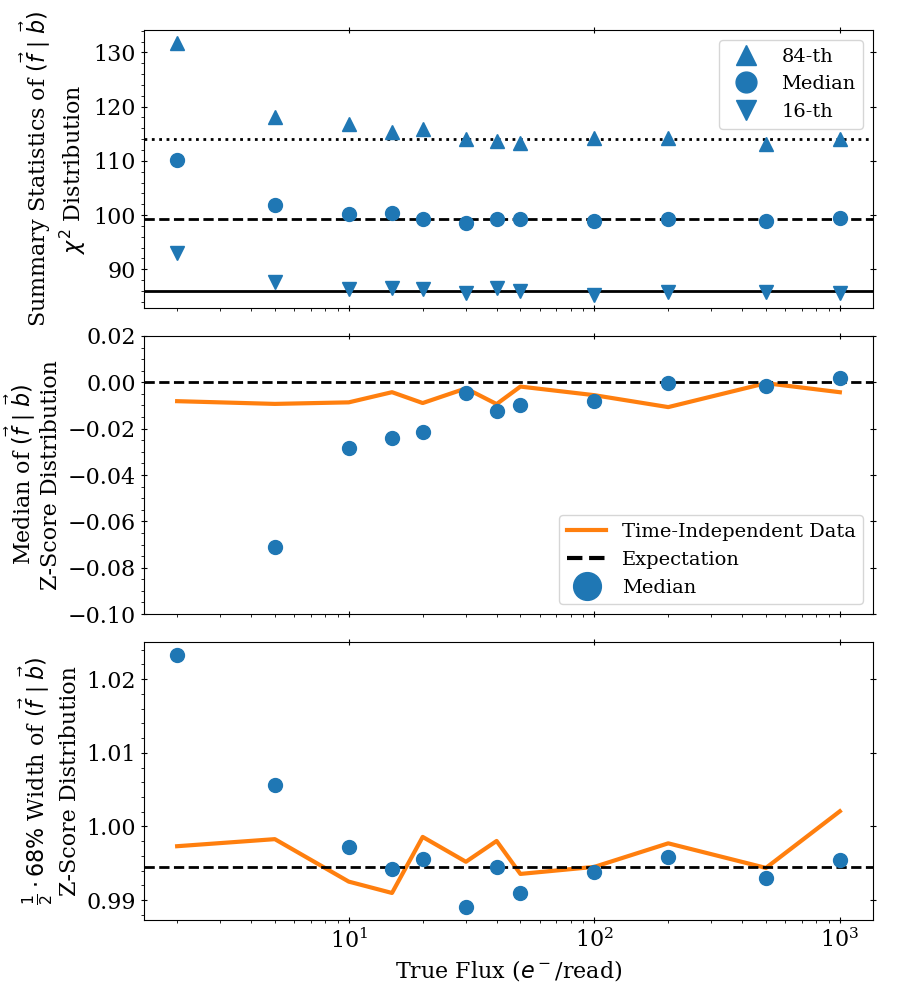}
    \caption{Summary of $(\vec f \mid \vec b)$ measurements from the same simulations in Figure~\ref{fig:Chi_distribution_of_b_vector}. For each realization, we correct the observed data using the best-fit $\vec b$ and then measure the best fit $(\vec f \mid \vec b)$ to simulate how real data would be processed. The top panel shows the 16-th, 50-th, and 84-th percentiles of the $\chi^2$ distribution of measured fluxes compared to the truth. The middle panel shows the median of the flux uncertainty-scaled residual distribution, with the orange line showing an equivalent analysis of time-independent data. The bottom panel estimates the width of the flux z-score distribution. This figure reveals that unbiased fluxes can be extracted using $\vec b$ measured from 100 pixels with flux levels $\geq30~e^-$\bl{/read}.}
    \label{fig:zscore_of_fluxes}
\end{figure}

%\begin{figure}[t]
%    \centering
%     \includegraphics[width=\linewidth]{figures/repeat_bvect_chi2.png}
%     \caption{The CDF for the $\chi^2$ of $\vec b$ vectors measured from 1000 iterations of fitting data with the same true $\vec b$ as shown in Figure~\ref{fig:synthetic_single_output_comp}. The different colored lines show the impact of changing flux levels (described in the text) as compared to the expected distribution (black line). Both the high and mid flux cases agree strongly with expectation, but the low flux case suggests our method overestimates the uncertainties of the $\vec b$ measurements.}
%     \label{fig:Chi_distribution_of_b_vector}
% \end{figure}

% \begin{figure}
%     \centering
%     \includegraphics[width=\linewidth]{figures/repeat_fit_f_given_bvect_zscore.png}
%     \caption{The CDF for the flux Z-scores measured from the same 1000 iterations of fitting data as Figure~\ref{fig:Chi_distribution_of_b_vector}. For each realization, we correct the observed data using the best-fit $\vec b$ and then measure the best fit $\vec f$ to simulate how real data would be processed. The different colored lines show the impact of changing flux levels (described in the text) which all have very good agreement with the expected distribution (black line).}
%     \label{fig:zscore_of_fluxes}
% \end{figure}

\subsection{Synthetic Observations over Different Flux Levels} \label{sec:fit_repeat_obs}

Next, we create 1000 realizations of data taken from the same configuration as in Section~\ref{sec:fit_single_obs} -- that is, using the same $n_\mathrm{pixels}$, $n_\mathrm{reads}$, read noise per pixel, and true $\vec b$ -- to quantify how well the measured $\vec b$ and $(\vec f \mid \vec b)$ vectors and uncertainties agree with the truth. We repeat this test across a range of flux levels: from read-noise-dominated cases to the photon-dominated regime. 

We first investigate our ability to measure $\vec b$ using data generated from different flux levels. The results of this test are shown in Figure~\ref{fig:Chi_distribution_of_b_vector}. The top panel shows the relationship between the median $\vec b$ uncertainty and the input true flux. Evidently, it takes a flux of $30$ to $40~e^-$\bl{/read} to reach a 10\% uncertainty in $\vec b$ when using 100 pixels with $n_\mathrm{reads} = 33$. We note that these tests used 100 synthetic pixels for all flux cases; it is theoretically possible to reach any desired SNR in $\vec b$ by increasing the number of pixels used in the joint fit, though this would significantly increase computation time. The middle panel next shows the 16-th, 50-th, and 84-th percentiles of the $\chi^2$ distribution of the measured $\vec b$ compared to the truth, with horizontal lines showing the expectation for those summary statistics. While most flux levels produce $\vec b$ distributions that agree with the expectation, a divergence begins to appear at low fluxes, suggesting that the chosen configuration requires a signal of $>10~e^-$\bl{/read} to measure trustworthy $\vec b$ vectors and uncertainties. At the highest flux level, we find that the $\vec b$ measurements slightly overestimate the uncertainties \bl{ based on the middle panel $\chi^2$ medians falling below the expectation. This size of this overestimate is} about 3\%, but too-large uncertainties are preferable to underestimated ones for most science applications. In the bottom panel, the medians of the $\vec b$ z-score distributions are close to the expectation of zero for most flux levels. For flux levels $\geq10~e^-$\bl{/read}, the bias is less than one percent of the measured $\vec b$ uncertainties; however, there are significant $\vec b$ biases for flux levels $<10~e^-$\bl{/read}. These results further emphasize the previous finding that jointly fitting 100 pixels requires fluxes $>10~e^-$\bl{/read} to have enough information to constrain $\vec b$. 

The most important question for this work, however, is whether the fluxes we measure are accurate. To that end, we take the best estimate of $\vec b$ for each realization and then use it to correct the data before measuring the final best fit flux, $(\vec f \mid \vec b_{\mathrm{best}})$. We then compare these fluxes and their uncertainties to the truth, as demonstrated by Figure~\ref{fig:zscore_of_fluxes}. The top panel shows that the flux $\chi^2$ distribution agrees with the expectation across most flux ranges. The middle panel shows that the extracted fluxes begin to become more biased than expected when the flux level is $<30~e^-$\bl{/read}. For the bottom two panels, we have included the results from analyzing comparable time-independent data (i.e., $\vec b = \vec 1$) using the \citet{Brandt_2024} constant flux model. We notice that both analyses tend to produce a slight negative bias in the z-score median, which is the result of a natural correlation between the extracted fluxes and their uncertainties: both the extracted flux and its uncertainty depend on a previous iteration's best guess of the flux, meaning they are fundamentally correlated such that we should not expect perfect agreement with the unit Gaussian. The bottom panel shows that the width of the flux z-score distribution agrees with expectations down to all but the lowest flux levels. 

These $(\vec f \mid \vec b)$ tests reveal that the limiting threshold is slightly higher than previously seen with the $\vec b$ results. To ensure unbiased $(\vec f \mid \vec b)$ in addition to trustworthy $\vec b$, we require the information content provided by 100 pixels with flux $\geq 30~e^-$\bl{/read}. We must emphasize that the results of this section do not imply that extracting low fluxes will always produce biased results; instead, it says that using the $\vec b$ measured from 100 low signal pixels can produce biased fluxes. In the next subsection, we will investigate the true flux bias when we are able to measure a precise $\vec b$.

\subsection{Synthetic Observations over Different $\vec b$ Amplitudes} \label{sec:fit_repeat_obs_b_amp}

\begin{figure}[t]
    \centering
    \includegraphics[width=1.0\linewidth]{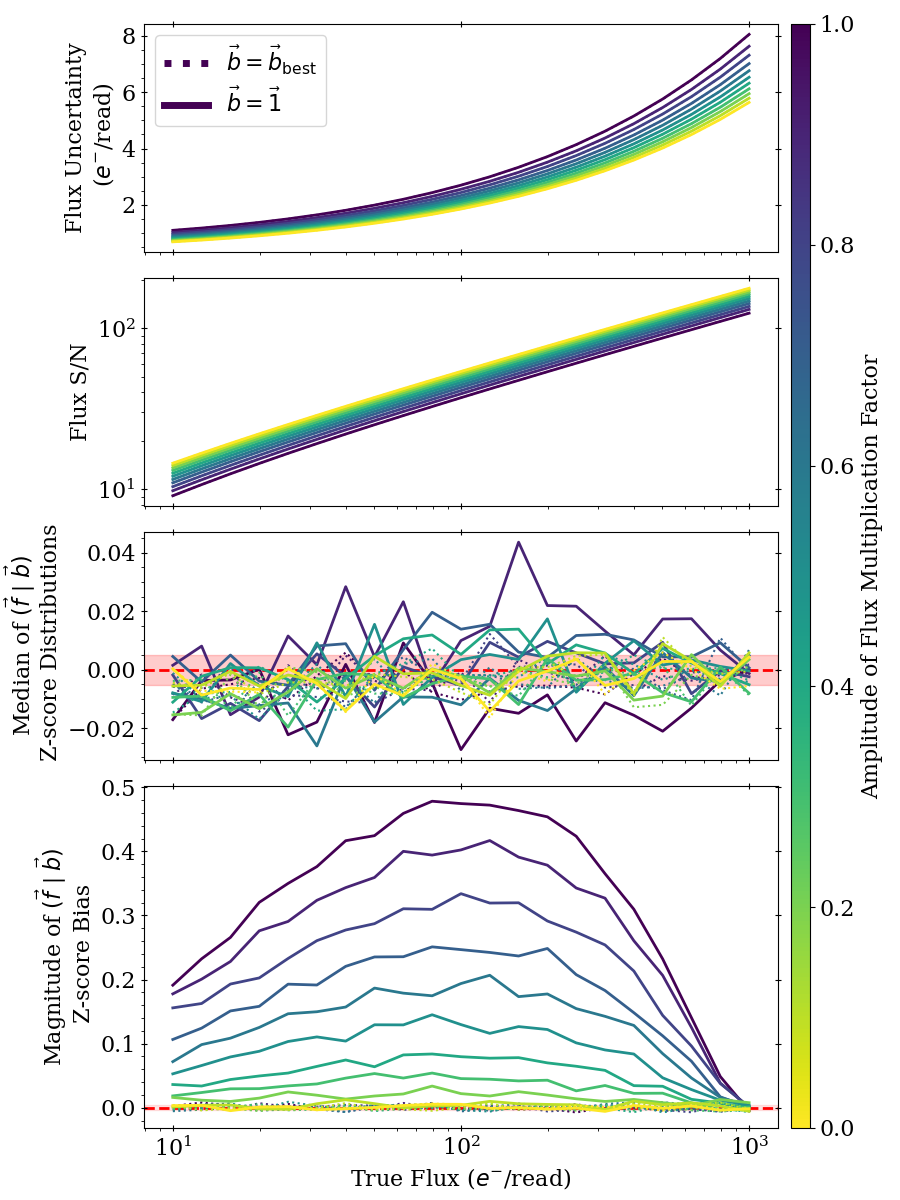}
    \caption{Quantification of flux uncertainty and bias as a function of true flux and $\vec b$ amplitudes from 0 (no change in flux versus time) to 1 (possible for fluxes to be 100\% diminished). Dotted lines correspond to fluxes extracted with the time-variable model, solid lines refer to fluxes extracted from the same data assuming time independence, and the color of each line corresponds to the $\vec b$ amplitude of those simulations. The top panel shows that the measured flux uncertainty increases with $b_\mathrm{amp}$ because fewer photons reach each pixel on average when $\vec b < \vec 1$. We note that the difference between the dashed and solid lines is $<1\%$ for all cases. \bl{The second-from-top panel shows the flux signal-to-noise, which is the true flux divided by the measured flux uncertainty.} The \bl{third-from-top} panel shows that the median of the uncertainty-scaled residual distributions marginalized over $\vec b$ realizations is broadly consistent with 0, though there is more scatter than expected (red shaded region) in the time-independent results. The bottom panel shows excess scatter in the median of the z-score distributions: $\sim 68\%$ of observations at a given $b_\mathrm{amp}$ are expected to show a bias with a magnitude equal to or smaller than the corresponding line. All of the dashed lines are very close to the expected red line, showing that the time-variable model produces the stochastically expected scatter in flux measurements. The solid lines show that fitting a constant flux to time varying data results in typical biases of $\sim 5\%$ of the flux uncertainty when $b_\mathrm{amp} = 0.3$ and up to $\sim50\%$ when $b_\mathrm{amp} = 1.0$. }
   \label{fig:synthetic_repeat_model_bias}
\end{figure}

\begin{figure}[t]
    \centering
    \includegraphics[width=1.0\linewidth]{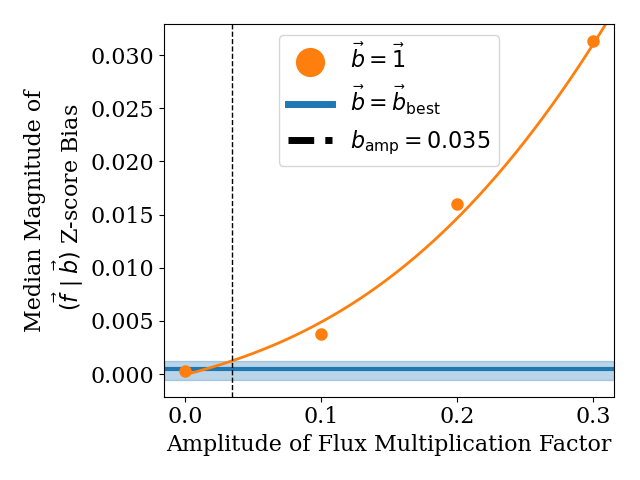}
    \caption{\bl{Measurement of the threshold for deciding the} optimal model for extracting fluxes. \bl{These results are based on} the bias magnitudes presented in the bottom panel of Figure~\ref{fig:synthetic_repeat_model_bias}. The y-axis is the median magnitude of the bias, and the x-axis is $b_\mathrm{amp}$. The blue line shows the median (over all fluxes and $b_\mathrm{amp}$) of the measured time-variable bias profiles, while the corresponding shaded region shows the 5 and 95\-th percentiles. The orange scatter points for the time-independent model show the per-$b_\mathrm{amp}$ median for each of the bottom panel Figure~\ref{fig:synthetic_repeat_model_bias} solid line profiles. The solid orange line is a simple 4-th order polynomial fit to the orange points, which has good agreement for $b_\mathrm{amp}\in [0,1]$. The black vertical line gives the $b_\mathrm{amp} = 0.035$ transition where the bias from the time-independent model has a $>95\%$ probability of being larger than the bias from the time-variable model.}
   \label{fig:synthetic_repeat_model_bias_transition}
\end{figure}

\begin{figure}[t]
    \centering
    \includegraphics[width=1.0\linewidth]{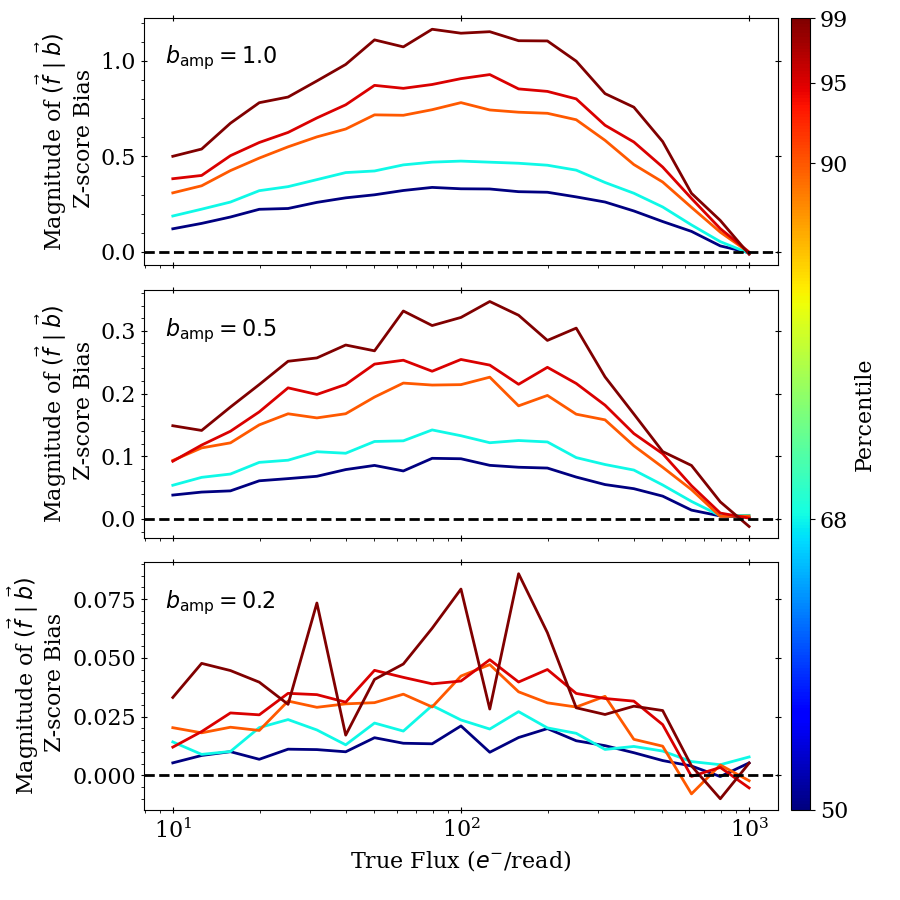}
    \caption{Magnitude of the flux bias at different percentile levels for $b_\mathrm{amp}\in[0.2,0.5,1.0]$. The cyan lines in these panels, corresponding to the 68-th percentile, are the same as the curves displayed in the bottom panel of Figure~\ref{fig:synthetic_repeat_model_bias}. The top panel reveals that the bias magnitude can be as large as $1.2\cdot \sigma_f$ for $b_\mathrm{amp} = 1.0$. The bottom panel suggests that we can be $\sim95\%$ confident that the time-independent model will produce a $<0.05\cdot \sigma_f$ magnitude bias when $b_\mathrm{amp} \leq 0.2$. }
   \label{fig:synthetic_repeat_model_bias_levels}
\end{figure}

Finally, we quantify the scale of the flux bias as a function of flux and amplitude in $\vec b$. That is, we define $ b_\mathrm{amp}$ such that a given true $\vec b_\mathrm{true}$ is created by $n_{\mathrm{reads}}-1$ draws from a uniform distribution with bounds from $1-b_\mathrm{amp}$ to 1. In this notation, $b_\mathrm{amp} = 0$ corresponds to the time independent case. We then draw 1000 realizations of $\vec b_\mathrm{true}$ vectors for $b_\mathrm{amp}\in[0,1]$ in steps of 0.1. We then follow the same process as described in Section~\ref{sec:fit_single_obs}: for each $\vec b_\mathrm{true}$ realization, we generate 100 pixels worth of data using a flux level of $1000~e^-$\bl{/read} and then use our linearized model to measure a high precision $\vec b_\mathrm{best}$. We next generate 100 examples of data across a range of fluxes with the same $\vec b_\mathrm{true}$, and then extract fluxes from those data using $(\vec f \mid \vec b = \vec 1)$ and $(\vec f \mid \vec b = \vec b_\mathrm{best})$. We then measure the flux uncertainties and residuals for each flux level for that $\vec b_\mathrm{true}$, giving one realization of the results shown in the top two panels of Figure~\ref{fig:synthetic_single_model_bias}. We then marginalize over the 1000 $\vec b_\mathrm{true}$ simulations for a given $b_\mathrm{amp}$, yielding the results in Figure~\ref{fig:synthetic_repeat_model_bias}. The different colored lines correspond to different $b_\mathrm{amp}$ levels, while dotted and solid lines represent the time variable and time independent models, respectively. Red lines and shaded regions show the expected values and scatters after drawing independent measurements from unit Gaussians using the same number of realizations as the simulated data. 

The top panel shows that the flux uncertainty measured by both models increases with $b_\mathrm{amp}$. This is because the number of photons that arrive at a pixel decreases for each increase in $b_\mathrm{amp}$, thereby reducing the data's constraining power. There is very little difference ($<1\%$) between the dotted and solid lines in this panel. The \bl{third} panel shows the median of $(\vec f \mid \vec b)$ z-score distributions, showing that marginalizing over all $\vec b$ realizations produces results that are fairly consistent with the expected 0. Here, we see that the time-variable model has smaller scatter than the time-independent model, especially as $b_\mathrm{amp}$ increases. We do notice that there is potentially a small bias ($\sim -0.5\%$ of the flux uncertainty) in the median z-score of the time-variable results as we move to lower fluxes. This is likely the same slight negative bias seen in the orange line in the middle panel of Figure~\ref{fig:zscore_of_fluxes}, suggesting that the new model is reaching the same fundamental limit as expected from time-independent data.

The bottom panel of Figure~\ref{fig:synthetic_repeat_model_bias} is a key result of this work as it gives the expected bias as a function of true flux and $b_\mathrm{amp}$. The plotted lines are calculated by taking the standard deviation of the 1000 realizations of the z-score medians for a given $b_\mathrm{amp}$, and then the expected scatter for this configuration of simulations is subtracted off (red line), resulting in the excess scatter in the z-score bias. All of the dotted lines for the time-variable model are very near the expected red line, suggesting the scatter we see in the time-variable results is explained by stochastic processes. The time-independent results, however, show that the measured scatter in the z-score bias is flux-dependent and grows to large values as $b_\mathrm{amp}$ increases. To provide an illustrative example, this plot says that a true flux near 100~$e^-$\bl{/read} can expect that, $68\%$ of the time, the bias will be as large as $0.5 \cdot\sigma_f$ in magnitude when $b_\mathrm{amp} = 1.0$. Of course, the precise bias sign and magnitude at each flux depend on the exact shape of $\vec b_\mathrm{true}$ for that observation (e.g., as shown in the \bl{third} panel of Figure~\ref{fig:synthetic_single_model_bias}). 

We can also use the results in the bottom panel of Figure~\ref{fig:synthetic_repeat_model_bias} to determine when it is better to use the time-variable or time-independent model, and this is demonstrated in Figure~\ref{fig:synthetic_repeat_model_bias_transition}. We have zoomed in on the $b_\mathrm{amp}\leq 0.3$ cases to highlight the transition between bias magnitude from the different models. The solid blue line is calculated by taking a median of the time-variable profiles across all flux levels and $b_\mathrm{amp}$, with the corresponding shaded regions giving the 5 and 95 percentiles. The blue line being slightly larger than 0 is evidence that the time-variable model incurs a small ($<0.1$\% of the flux uncertainty) cost associated with fitting $\vec b$ and then extracting $(\vec f \mid \vec b = \vec b_\mathrm{best})$. The orange points are calculated by taking medians over the flux levels for the solid profiles in the bottom panel of Figure~\ref{fig:synthetic_repeat_model_bias}, giving average bias magnitude per $b_\mathrm{amp}$. The orange solid line is the result of fitting a simple 4-th order polynomial to the orange points for all $b_\mathrm{amp}$, which does a good job of describing the measurements for $b_\mathrm{amp}\in[0,1]$. \bl{Tests with other polynomial orders -- both higher and lower than 4 -- did a visually worse job of describing the orange data points in the $b_\mathrm{amp} \leq 0.3$ region or produced unphysical interpolated values (e.g., relatively large negative values in the $b_\mathrm{amp} \leq 0.1$ region).} We find that the solid orange line becomes greater than the top of the blue shaded region at $b_\mathrm{amp} = 0.035$. This implies that there is a $>95\%$ probability that the time-independent model contributes more systematic error to the measurements than the time-variable model when the time-variable signal is $>3.5\%$. We suggest that $b_\mathrm{amp}>3.5\%$ is a useful threshold for deciding when the time-variable model is preferred. \bl{Using linear interpolation of the orange points instead of the 4-th order polynomial returns an intersection at $b_\mathrm{amp} = 0.028$, suggesting a more conservative choice would be to prefer the time-variable model when the time-variability is $>2.8\%$.}

We explore different percentiles of the bias magnitude distributions \bl{of the time-dependent model} for $b_\mathrm{amp}\in[0.2,0.5,1.0]$ in Figure~\ref{fig:synthetic_repeat_model_bias_levels}. The cyan lines in this figure are the same as the solid curves plotted in the bottom panel of Figure~\ref{fig:synthetic_repeat_model_bias}. From the top panel, we see that the bias magnitudes can become as large as 120\% of the flux uncertainty. A key takeaway comes from the bottom panel, which shows that we can be fairly confident -- 95\% of the time -- that the time-independent model will produce flux-dependent biases that are smaller than 5\% of the flux uncertainty when $b_\mathrm{amp}\leq 0.2$. These metrics provide another possible threshold for determining when to transition between the time-variable and time-dependent models. 

We emphasize that the flux levels, read noise, and number of reads used in our synthetic tests are characteristic of real APOGEE observations. As a result, Figures~\ref{fig:synthetic_repeat_model_bias} to Figures~\ref{fig:synthetic_repeat_model_bias_levels} provide useful metrics to assess the impact of ignoring time-variability in data. Using independent estimates of the APOGEE flux change as a function of time -- for example, from a guider camera on the telescope -- one could estimate the magnitude of the flux bias to decide whether to extract using the time-independent or time-dependent model. 

\begin{figure}[t]
    \centering
    \includegraphics[width=\linewidth]{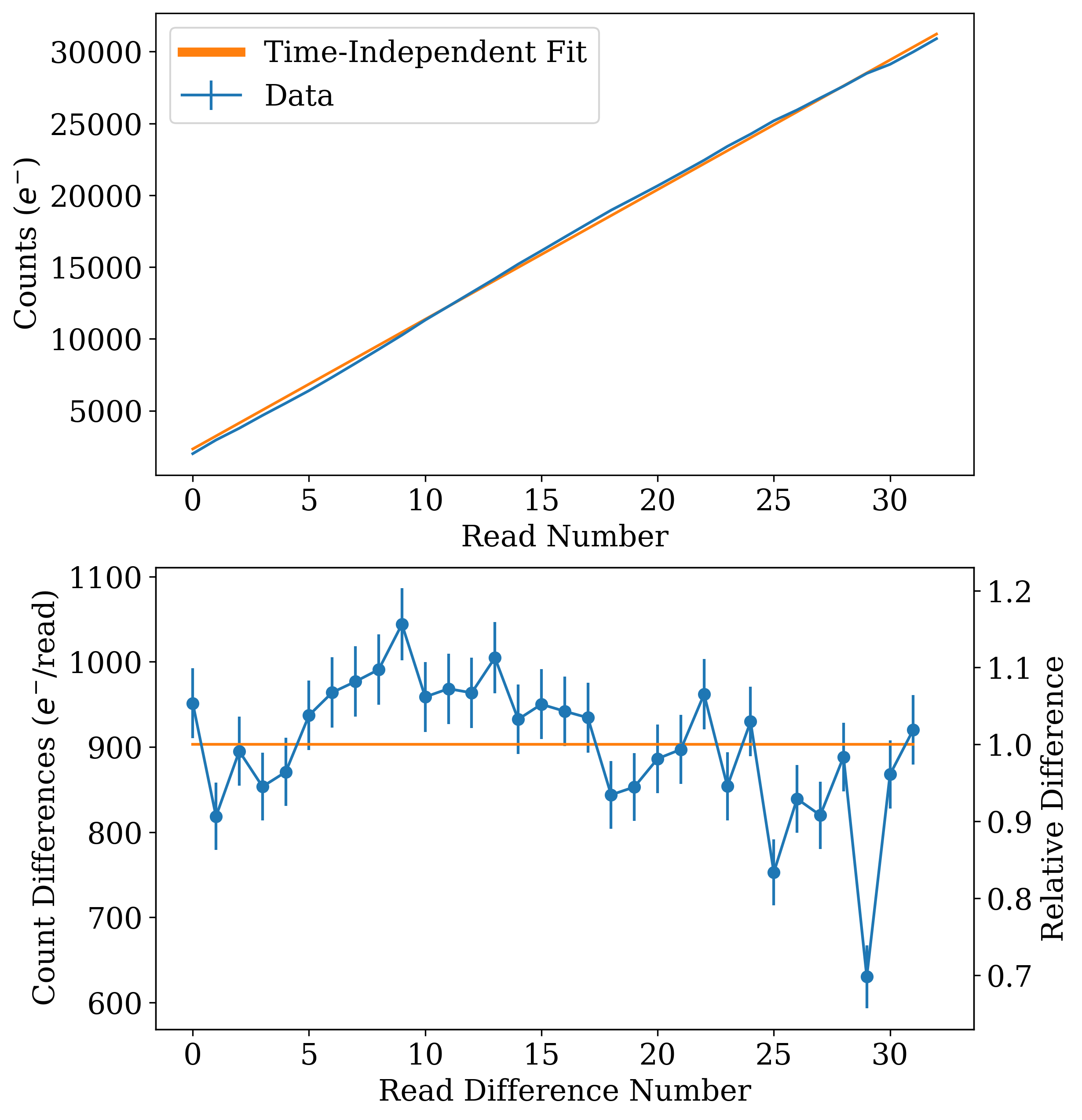}
    \caption{\textbf{Top:} An example 33-read ramp of counts versus read time using real science data from the APOGEE spectrograph (blue line) compared to the best-fit constant flux (orange line).
    \textbf{Bottom:} Count differences revealing time fluctuations in the data that are not well-described by a single constant flux. \bl{The right axis shows the relative count differences after dividing by the best-fit constant flux.}}
    \label{fig: Counts vs Read Times using real data.}
\end{figure}

\section{Validation with Real Data} \label{sec:apogee_data}

% \begin{figure}
%     \centering
%     \vspace{1em}
%     \includegraphics[width=\linewidth]{figures/b_vector_linear.png}
%     \caption{Example of the b vector in APOGEE data fitting. The b vector fluctuates significantly.}
%     \label{fig: b_vector}
% \end{figure}

\begin{figure}
    \centering
    \includegraphics[width=\linewidth]{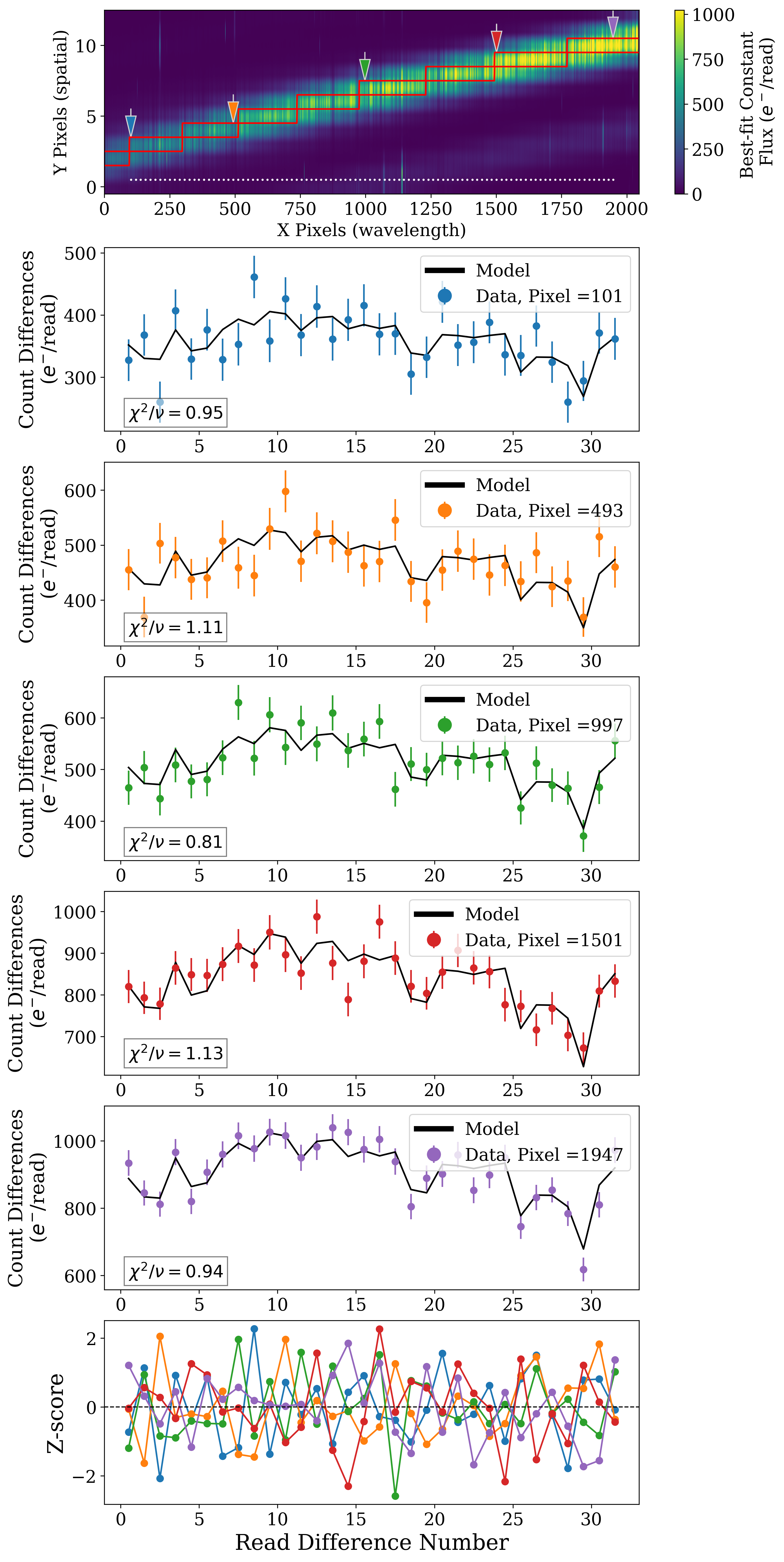}
    \caption{Comparison of the best-fit model with real APOGEE data across many pixels spanning $\sim 600$~Angstroms in wavelength. The pixels come from one APOGEE detector \bl{(chip R)} and all lie on the trace of the same star \bl{with the top panel showing the star's 2D spectrum (red bounding lines showing the pixels nearest to the trace center). White dots near the bottom of the top panel show the X positions of the 100 pixels used in measuring $\vec b_\mathrm{best}$, and colored arrows point to the 5 pixels highlighted in the following panels of this figure}. The black line shows the model prediction in that pixel, while the colored points show the count differences in each pixel. The $\chi^2$ comparison between model and data is shown in the bottom left corner of each panel, scaled by the relevant degrees of freedom. The last panel displays the uncertainty-scaled residuals, showing that the model does a good job of describing the data.}
    \label{fig: combined_all_fibers}
\end{figure}

\begin{figure}
    \centering     
    \includegraphics[trim={0cm 0cm 0cm 0cm},clip,width=\linewidth]{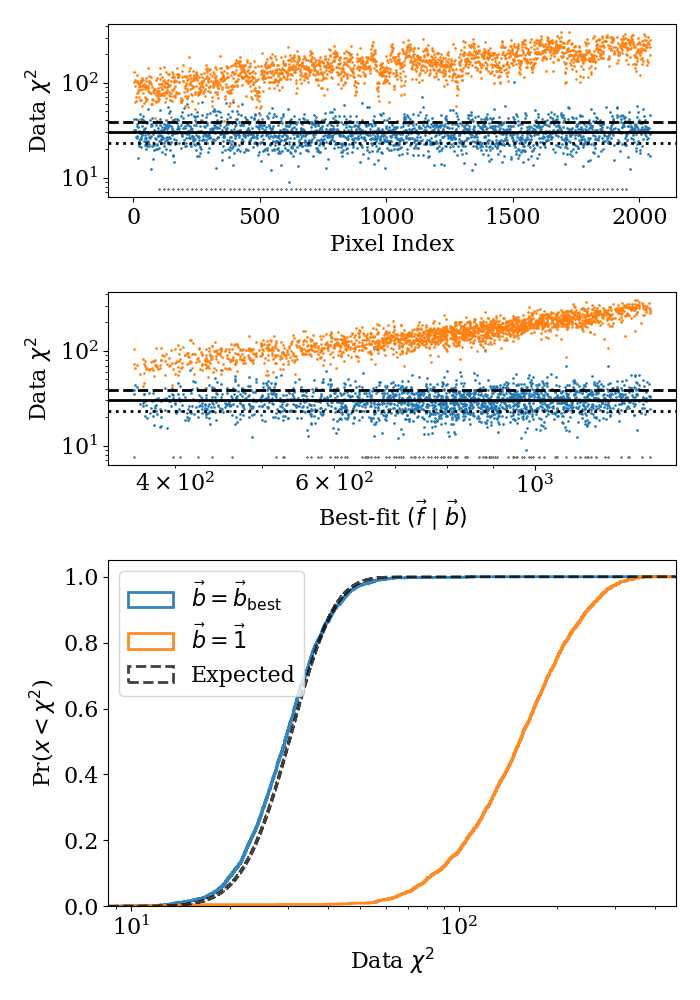}
    \caption{Comparison of the data residual $\chi^2$ distributions for 2048 pixels of the same star used in the analysis of Figure~\ref{fig: combined_all_fibers}. The two upper panels show the data $\chi^2$ per pixel as a function of pixel index (i.e. wavelength) and flux, respectively, while the bottom panel shows the empirical CDFs. The black horizontal lines in the upper panels show the 16-th, 50-th, and 84-th percentiles of the expected $\chi^2_\nu$ distribution. Small grey points along the bottom edges of the top two panels show the pixel indices and fluxes for the 100 equally-spaced pixels used in the $\vec b$ fit. The orange points and distribution comes from using the best-fit time-constant flux as defined by the \citep{Brandt_2024} model, while the blue points and distribution show the results after accounting for time variation using the new model of this work. The time-varying model almost perfectly matches the expected distribution, while the time-independent model does not. These distributions reveal that the time constant model requires the data uncertainties to be inflated by a factor of 2.27 to have the expected agreement, while the time varying model requires multiplication by a factor of 0.99.}
    \label{fig: chi with real data}
\end{figure}

\begin{figure}
    \centering
    \includegraphics[width=\linewidth]{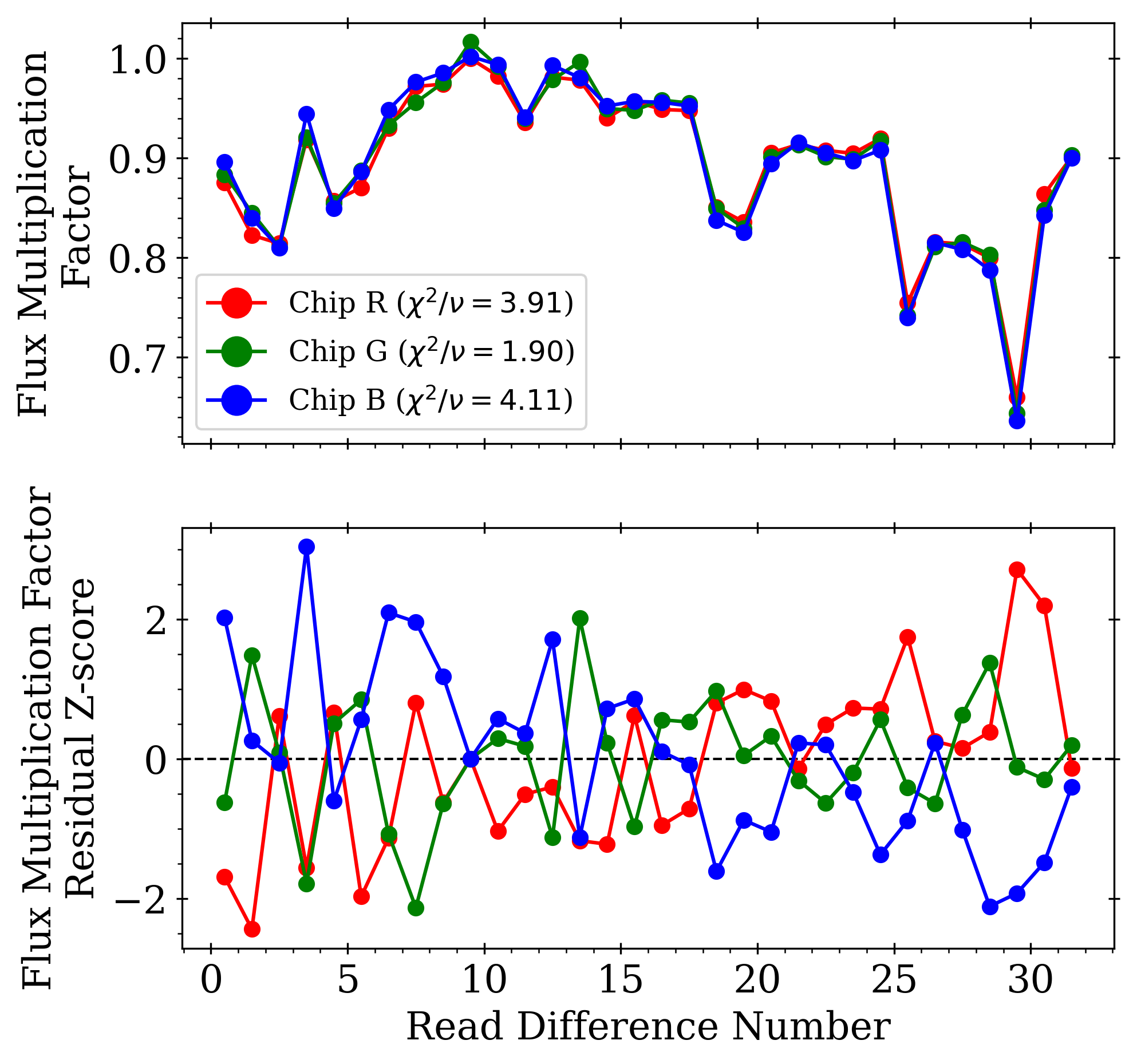}
    \caption{\textbf{Top:} Examples of $\vec b$ vectors (re-scaled by their median) for different APOGEE chips/detectors covering different wavelength ranges for the same exposure and star as in Figures~\ref{fig: combined_all_fibers} and \ref{fig: chi with real data}. Legends give the scaled $\chi^2$ comparison between each chip's $\vec b$ and an inverse-variance weighted-average \bl{of $\vec b$ calculated from all three chips}. \textbf{Bottom:} Uncertainty-scaled residuals of the $\vec b$ vectors in the top panel compared to an inverse-variance weighted-average. Even though the data on the different chips were fit independently, the results are extremely similar, indicating that our assumption of a wavelength-independent $\vec b$ is broadly supported by the data.}
    \label{fig: b_vector_compare}
\end{figure}

We also validate our model using real data from the APOGEE spectrograph at Apache Point Observatory, using observations that were part of the Sloan Digital Sky Survey V \citep[SDSS-V;][]{Kollmeier_2025}. 

\subsection{\bl{Time Variable Signals in Real Science Observations}}
\label{sec:real_data_science}

A flux ramp from an example real \bl{science} pixel is shown in the top of Figure~\ref{fig: Counts vs Read Times using real data.}, with the best-fit constant flux shown in orange; this is a pixel on the trace of a bright star, away from sky emission lines, at a wavelength of $\sim 16928.1$ \r{A}. After taking the difference between subsequent reads, the bottom panel reveals that the flux is varying by amounts larger than the uncertainties can explain, suggesting that a time-constant flux is not a good model of the data.

Using our time-varying model, we fit the data from 100 pixels simultaneously. These pixels are taken by following the maximum of the trace for one relatively bright star across a single detector. That is, these pixels correspond to the same star and position on the sky but cover a wide range in wavelength ($\sim 600$~\r{A}). A comparison between the resulting best fit model and count differences is shown in Figure~\ref{fig: combined_all_fibers}, with the different panels highlighting different pixels (i.e., wavelengths). The uncertainty-scaled residuals in the bottom panel show that the new model does a good job of describing the data. 

As with the synthetic data, we then use the best-fit $\vec b$ to correct the data before measuring a new best-fit flux in each pixel, $(\vec f \mid \vec b = \vec b_\mathrm{best})$, including pixels that were not used in the $\vec b$ measurement\bl{. That is, we use all} 2048 pixels along the trace \bl{center (all pixels inside of the red bounded region in the top panel in  Figure}~\ref{fig: combined_all_fibers}\bl{) instead of only the 100 pixels used to measure $\vec b_\mathrm{best}$}. We then compare the data $\chi^2$ we obtain with our new measurements versus the constant-flux results, yielding the distributions in Figure~\ref{fig: chi with real data}. Here, we see that the time-varying model has removed the disagreement seen between the expected distribution and the time-independent model. The upper two panels also reveal that the time-variable model does not show any noticeable trends in data $\chi^2$ as a function of flux and wavelength -- which is a huge improvement over the time-independent model -- lending credence to our assumption that the majority of the temporal variation is achromatic. 

We further test the wavelength independence assumption by analyzing data from two additional detectors that were observing the same targets at the exact same times, just at different wavelengths. We will refer to the different chips using ``red'', ``green'', and ``blue'' \bl{-- labeled R, G, and B --} based on their relative positions in wavelength. We repeat our analysis using the trace of the same star, but now using data from the other chips analyzed completely independently. The best-fit $\vec b$ vectors for each of the three chips are shown in Figure~\ref{fig: b_vector_compare}. The pixels now span almost 2000 \r{A} in wavelength, and yet the results from three separate $\sim 600$~\r{A} regions agree \bl{quite closely} with \bl{each other. However, we do note that the chips show some statistical differences when their uncertainties are accounted for, as highlighted by the reduced $\chi^2$ legend values}. Some of the differences seen between the chips in their residuals compared to their weighted average (bottom panel) might be due to different nonlinearity responses that have not been removed from the data, varying persistence properties, chromatic differences in point spread functions, or a high order wavelength-dependence in $\vec b$. 

The APO Sloan Telescope covers a three degree wide field of view.  When we explore the $\vec b$ that we measure from fibers/stellar traces that correspond to positions that are nearby on the sky, we find a general agreement. We are still in the early stages of understanding how these results change across the focal plane.  Small, spatially-coherent differences may be telling us about the individual atmospheric differences each fiber sees along its line-of-sight, or they may be indicative of a variation in telescope-delivered image quality across the focal plane. We are actively exploring how best to measure and correct the observed data as a function of position on the focal plane, source brightness, and position on detector. Implementing our time-variable model using an entire detector's/night's worth of data will be the focus of future work. 

\begin{figure}
    \centering
    \includegraphics[width=\linewidth]{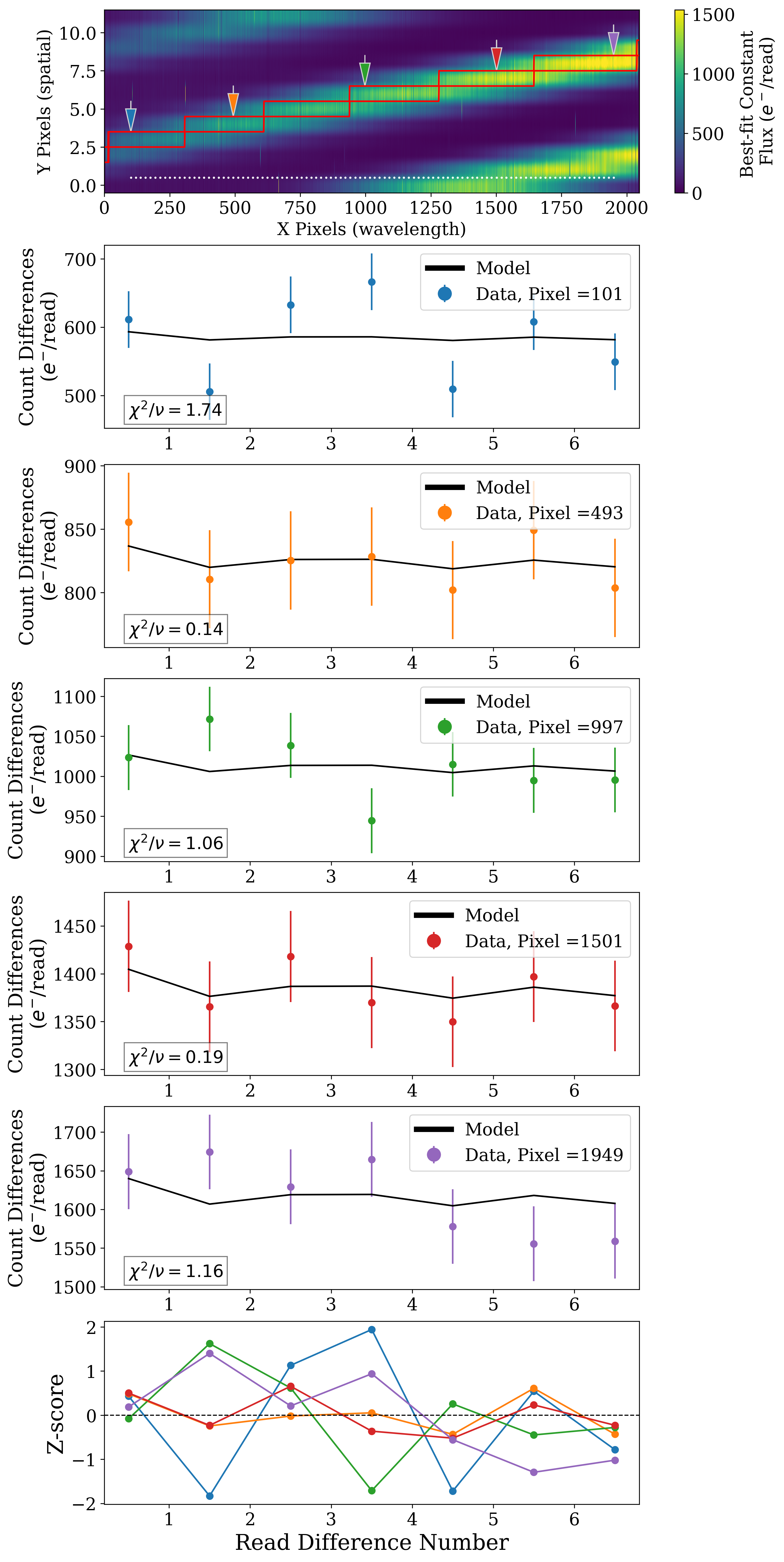}
    \caption{Same as Figure~\ref{fig: combined_all_fibers}, but now for APOGEE quartz lamp calibration data \bl{(chip R)}. The source in this case is inside of the telescope's dome, so we do not expect to see a significant variation in time from the atmosphere. Indeed, we recover a $\vec b$ that is consistent with a time-stable source \bl{for data on this chip}.}
    \label{fig: model with calibration data}
\end{figure}

\begin{figure}
    \centering     
    \includegraphics[trim={0cm 0cm 0cm 0cm},clip,width=\linewidth]{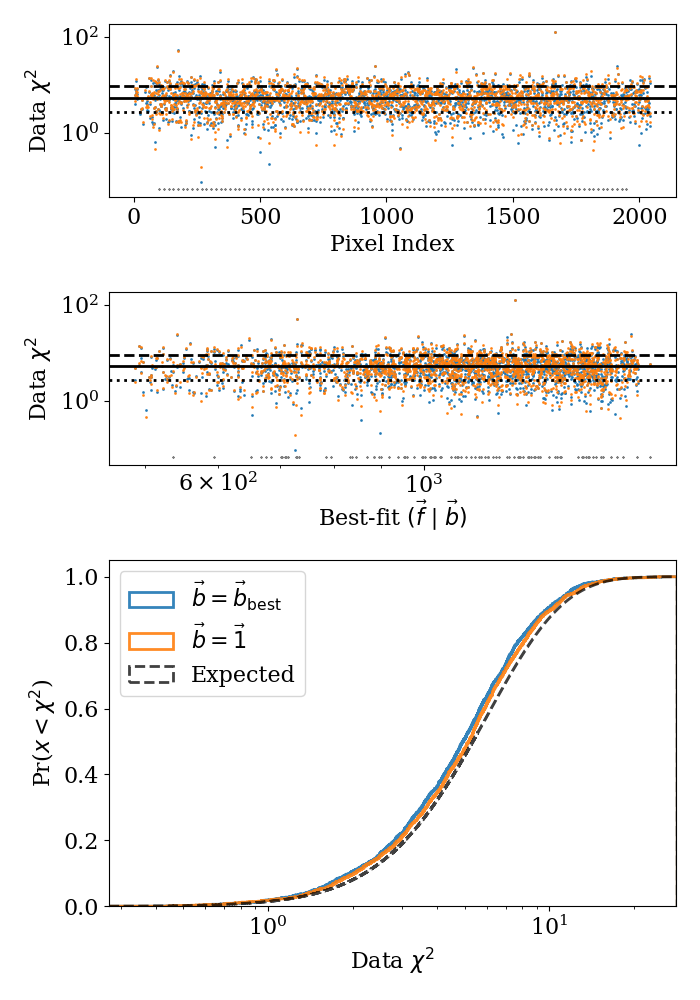}
    \caption{\bl{Same as Figure}~\ref{fig: chi with real data}\bl{, but using the calibration data in Figure}~\ref{fig: model with calibration data}\bl{. As expected, there is little difference between the time-independent and time-dependent results for the calibration data where time variation should be minimal.}}
    \label{fig: chi with real calibration data}
\end{figure}

% \begin{figure}
%     \centering
%     \includegraphics[width=\linewidth]{figures/calibration_linear_b_vect.png}
%     \caption{Best-fit $\vec b$ (re-scaled by their median) for the calibration data shown in Figure~\ref{fig: model with calibration data}. The red-marked point is the read time that was held fixed (to exactly 1.0) during fitting, as discussed in Section~\ref{sec:statistics}; it now has a value $>1$ because of the median re-scaling of $\vec b$. Taking the full posterior covariance matrix of $\vec b$ into account, we find that the flux multiplication factor is consistent with a vector of ones, agreeing with the expectation that these data do not exhibit time evolution.}
%     \label{fig:calibration_b_vect}
% \end{figure}

\begin{figure}
    \centering
    \includegraphics[width=\linewidth]{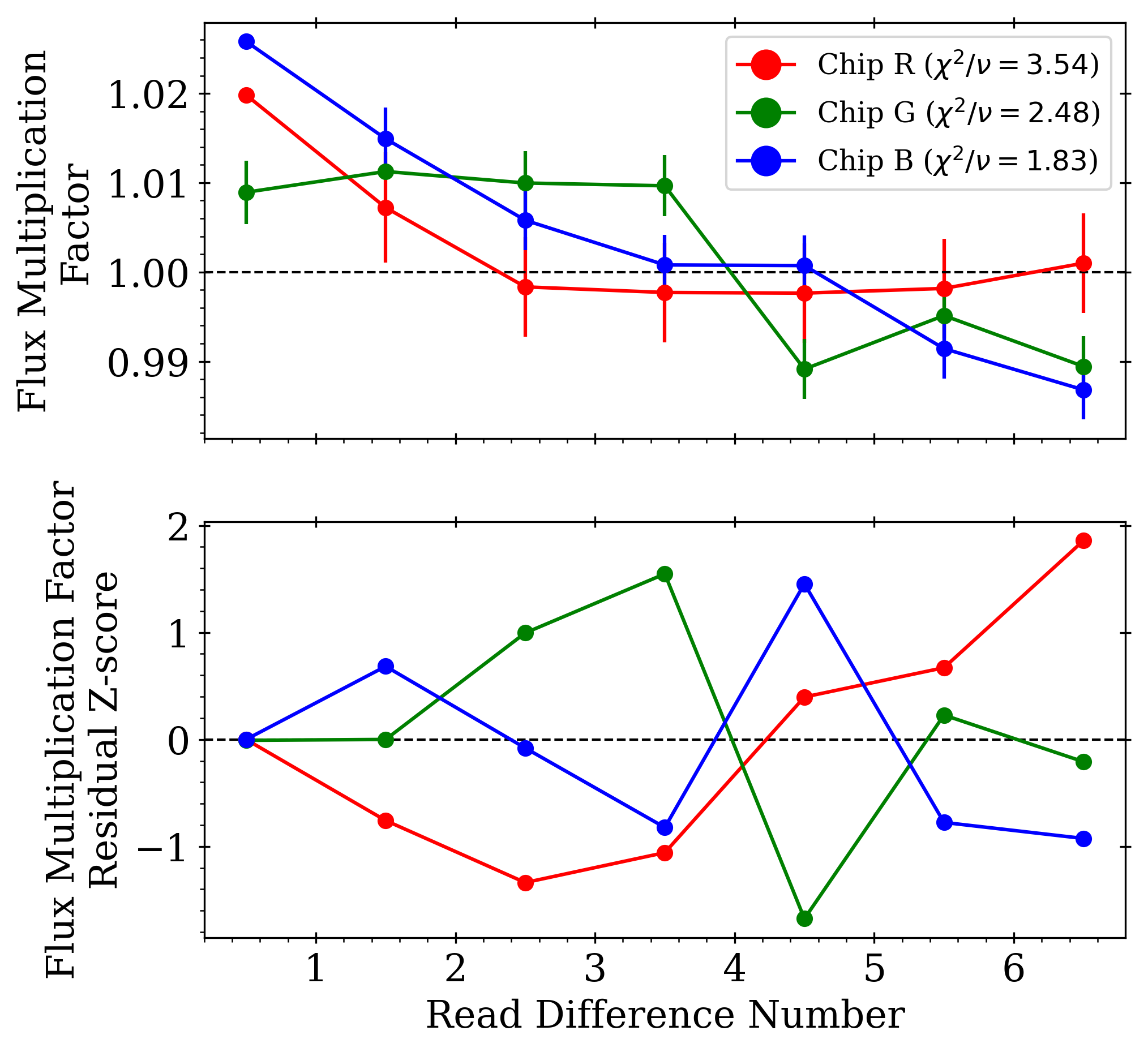}
    \caption{\bl{Same as Figure}~\ref{fig: b_vector_compare}\bl{, but using the calibration data in Figure}~\ref{fig: model with calibration data}\bl{. The reduced $\chi^2$ values presented in the legend again compare each chip's $\vec b$ to a weighted average $\vec b$, not to the expected vector of ones. When we instead compare to the expected $\vec b=\vec 1$ case, we measure reduced $\chi^2$ values of 0.78, 15.8, and 15.9 for the R, G, and B chips respectively. These results suggest the R chip is consistent with the time-independent expectation, but the other chips are not. For both the G and B chips (and R to a lesser extent), we see a decrease in $\vec b$ over time, which is what we would expect from nonlinearity effects.} }
    \label{fig:calibration_b_vect}
\end{figure}

\subsection{\bl{Null Tests with Telescope Calibration Data}} \label{sec:real_data_telescope_calibration}

Next, we test our model on telescope calibration exposures that we expect to be time-independent. The particular exposure we use is of quartz lamp light that is reflected off the interior of the telescope dome. This means the traces are bright, all fibers are evenly illuminated, and the measurements should not exhibit atmosphere-related time variability. There may be some subtle sources of flux variability -- such as the lamps heating up over the course of use, persistence, or incompletely-removed nonlinearity effects -- but we expect these to be small compared to the changing sky seen with the real science data. After repeating the same analysis using 100 pixels across one detector for a single fiber's trace, we get the results in Figure~\ref{fig: model with calibration data}. As expected, the $\vec b$ is quite flat\bl{. We find that both the time-independent and time-dependent models do a good job of describing the data with no obvious bias as a function of flux or wavelength, as demonstrated in Figure}~\ref{fig: chi with real calibration data}\bl{. The data $\chi^2$ distributions suggest that the data uncertainties might be slightly overestimated by a factor of $2-3$\%, providing potential evidence that the current read noise estimates are slightly too large.}

\bl{We again compare the $\vec b$ measured from the different chips and assess how similar they are to each other as well as to the expected $\vec b=\vec 1$. These results are shown in Figure}~\ref{fig:calibration_b_vect}\bl{. We find that all the $\vec b$ elements from the different chips are within a few percent of the expected one vector. The chips also all generally show a decrease in $\vec b$ elements as a function of time, suggesting that the time-dependent signal we measure in this case is likely due to nonlinearity effects. We note that the reduced $\chi^2$ values in the upper panel legend are the results when comparing each chip's $\vec b$ to a weighted-average $\vec b$. As with the previous science observations, we see some disagreement between the chips when we account for the $\vec b$ uncertainties, which is most likely a result of different nonlinearity properties. Statistically comparing each chip's $\vec b$ to the expected $\vec b = \vec 1$, we measure reduced $\chi^2$ values of 0.78, 15.8, and 15.9 for the R, G, and B chips respectively. While chip R is consistent with a time-stable source, the other chips are not, potentially indicating that $\vec b$ is truly capturing nonlinearity behavior and picking up differences between the chips. These nonlinearity differences may also explain the nature of the small disparity between the $\vec b$ measured from real science data in Figure}~\ref{fig: b_vector_compare}\bl{. In the future when we have access to nonlinearity corrections for these data, we would expect to see the chip differences reduce and for the $\vec b$ measurements to be more consistent with a vector of ones.}

\subsection{\bl{Comparison of Time Variable Signal with Independent Sky Measurements}}
\label{sec:real_data_comp_with_guider}

Finally, we compare the $\vec b$ we measure to independent estimates of the sky's properties as determined by the Sloan Telescope's guide camera system. This system consists of six CCDs equipped with SDSS $r$ band filters mounted around the periphery of the telescope's field of view \citep{sdss_fps}.  All guide cameras are exposed in a continuous loop during survey observations.  Point sources detected in guide frames are cross-matched to Gaia DR2 \citep{gaia_dr2} objects with $G<18$~mag.  Gaia astrometry is used to determine telescope pointing corrections during science exposures, but the guide frames additionally track changes in flux (transparency) and seeing (FWHM) for each unique source detected.

Variations in transparency are generally associated with changes in cloud cover, and we expect transparency trends measured by the guide system to also be present in the spectra.  In addition to transparency, atmospheric seeing can vary on timescales shorter than a typical spectrograph exposure.  The SDSS-V guider reports a Gaussian FWHM for each guide exposure. Seeing can significantly affect the flux received at the spectrograph in fiber-fed spectroscopy.  APOGEE fibers subtend a 3 arcsecond diameter circle on sky at the Sloan Foundation Telescope, so when seeing is poor (e.g. $>$ 3 arcsec FWHM), significant amounts of the available photons are landing outside the fiber's collecting area.

\bl{Given measurements for transparency and seeing, we can roughly model the flux-in-fiber change based on $n_g$ guide camera images taken during an APOGEE exposure.  We choose a 2D Gaussian model for the telescope's PSF during guider exposure $i$ (with $i\in[1,\dots,n_g]$)}
\begin{equation}
g(x, y, \sigma_i) = \frac{1}{2\pi\sigma_i^2}\exp{\left( - \left[ \frac{x^2+y^2}{2\sigma_i^2}\right] \right)},
\end{equation}
\bl{where the Gaussian width is informed by the guider measured FWHM in arcseconds}
\begin{equation}
\sigma_i= \frac{\mathrm{FWHM_i}}{2\sqrt{2 \ln 2}}.
\end{equation}
\bl{For each guide image, we analytically integrate the PSF profile over the surface of the circular 3 arcsecond diameter fiber, and then multiply by the guider measured transparency $t_i$ to estimate the flux entering the fiber}
\begin{equation}
f_i = t_i\int_{0}^{2\pi}\int_{0}^{1.5''}g(r\cos\theta,r\sin\theta, \sigma_i)r \,dr\,d\theta.
\end{equation}
\bl{Finally, we normalize this flux sequence by the maximum computed flux during the APOGEE exposure}
\begin{equation}
f_{i,\mathrm{norm}} = \frac{f_i}{\max \{f_1, f_2, \dots, f_{n_g}\} }. 
\end{equation}
Telescope pointing errors (or guide jitter) \bl{can} also be a source of time-variable flux entering the fiber.  In the \bl{examples we presented next, the} guide error is stable and small, and so should not contribute toward a time-variable flux signal in the spectra.

\begin{figure}[t]
    \centering
    \includegraphics[trim={0cm 0cm 0cm 1.3cm},clip,width=\linewidth]{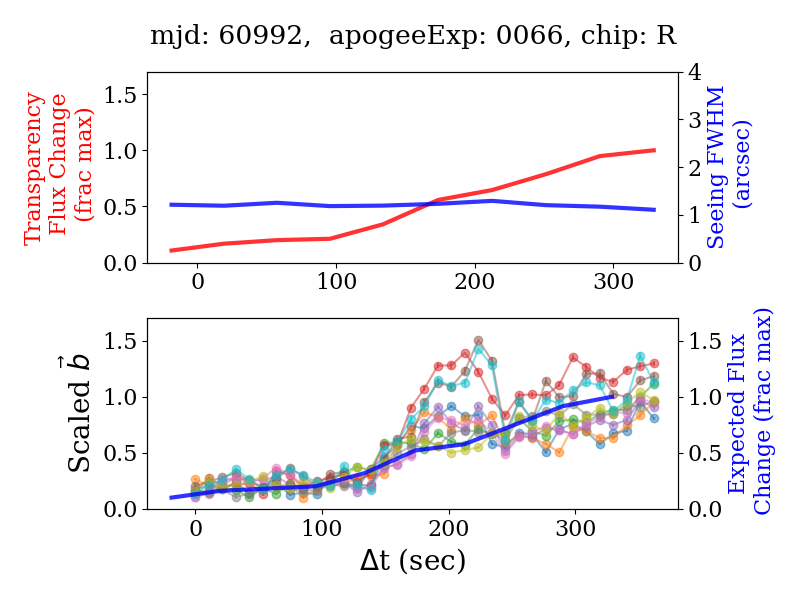}
    \caption{Comparison of $\vec b$ measured from the ten brightest stars in an APOGEE exposure (colored lines with scatter points, left axis of bottom panel) with guider camera measurements of transparency (red line, left axis of top panel), seeing (blue line, right axis of top panel), and total estimated change in flux (dark blue line, right axis of bottom panel). While the seeing is relatively stable over the APOGEE exposure, the transparency increases from beginning to end and matches the shape of the measured $\vec b$ vectors, suggesting that a cloud moving out of the field of view is driving the time-varying component of the fluxes. }
    \label{fig:guiderComp_trans_change}
\end{figure}

% From guider data taken during an APOGEE exposure we can look for variations in sky transparency, seeing, and/or telescope pointing errors. Changes in any one of these quantities during an APOGEE exposure can impact the amount of flux reaching the detector versus time and manifest as signal in $\vec b$: (1) transparency variation will typically track changes in cloud cover; (2) an APOGEE fiber covers a 3 arcsecond diameter patch of sky, so as the seeing point spread function size increases, less light will enter the fiber; (3) unstable guiding behavior could move flux in and out of the fiber throughout an APOGEE exposure. The typical exposure time for the SDSS-V guide system is $15$~seconds with $\sim 37$~seconds between each exposure, compared to the $\sim 10.6$~seconds between readouts of the APOGEE detectors which are taken immediately one after another. Differences in timing and wavelength coverage mean that we do not expect perfect agreement between the guider and APOGEE measurements, so we instead focus on qualitative comparisons.

We start by considering a case where the guider suggests that cloud cover is dispersing during the exposure while the seeing remains relatively constant. In this case, we measure $\vec b$ for ten of the brightest stars, each fit independently, using 100 pixels along each star's trace on a single detector. The results are shown in Figure~\ref{fig:guiderComp_trans_change} with the colored lines in the bottom panel showing the best fit $\vec b$ measurements, rescaled to have the closest agreement with the guider-estimated flux change. The top panel shows the median guider-measured transparency change across all sources in all guide chips (red line, left axis) in addition to the seeing FWHM (blue line, right axis). The right axis of the bottom panel displays the estimated flux change after combining the effects of changing transparency and seeing.  The agreement in shape between the transparency and the $\vec b$ from the different stars suggests that the change in flux as a function of time is largely dominated by transparency changes for this exposure. 

\begin{figure}[t]
    \centering
    \includegraphics[trim={0cm 0cm 0cm 1.3cm},clip,width=\linewidth]{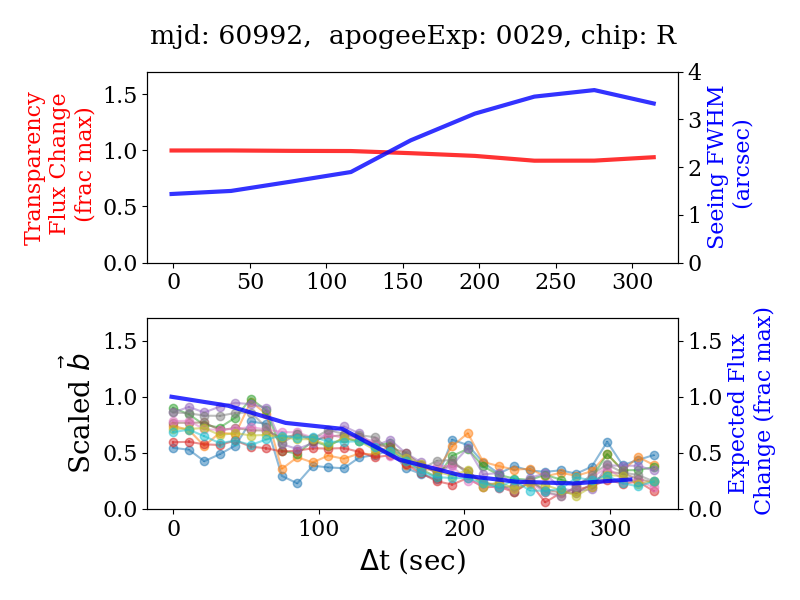}
    \caption{Same as Figure~\ref{fig:guiderComp_trans_change} but using a different exposure for a case where the seeing changes while the transparency is relatively stable. As the seeing increases from $\sim 1.5$~arcsec to $\sim 3.5$~arcsec, less of each star's light is being collected by the fiber compared to the beginning, driving the change in $\vec b$ as shown by the the bottom panel.}
    \label{fig:guiderComp_seeing_change}
\end{figure}

\begin{figure}[t]
    \centering
    \includegraphics[trim={0cm 0cm 0cm 1.3cm},clip,width=\linewidth]{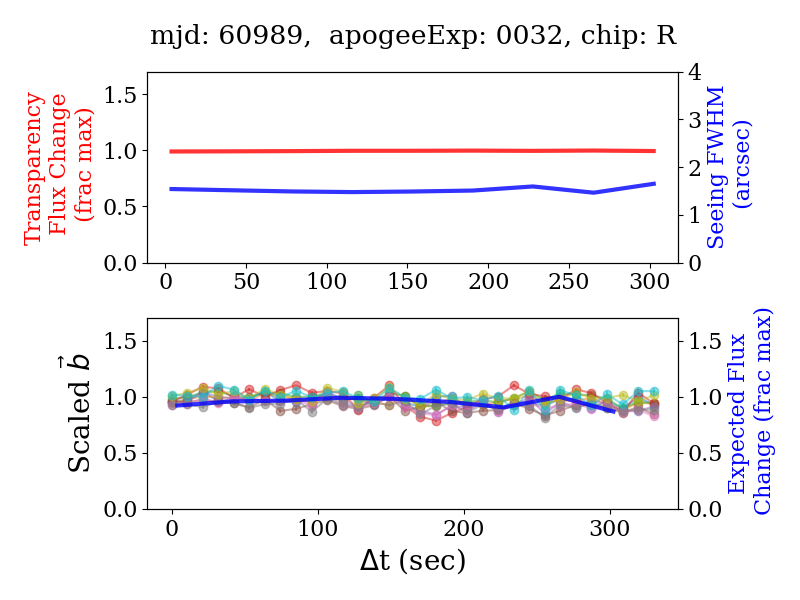}
    \caption{Same as Figure~\ref{fig:guiderComp_trans_change} but using a different exposure for a case where the transparency and seeing are both relatively stable. All the $\vec b$ measurements are quite close to 1.0, agreeing with a largely stable atmosphere.}
    \label{fig:guiderComp_no_change}
\end{figure}

We then repeat this analysis for two additional APOGEE exposures. The next example, shown in Figure~\ref{fig:guiderComp_seeing_change}, is a case where the seeing is changing but the transparency is stable. These results show the expected anti-correlation between the $\vec b$ measurements and seeing: as the seeing increases, less light is entering the fiber compared to earlier in the exposure. The final example, shown in Figure~\ref{fig:guiderComp_no_change}, is where the guider predicts that the atmosphere is stable and the measured $\vec b$ vectors largely agree. 

With these real data tests, we have shown that the model is flexible enough to process images under a variety of conditions: large versus small time variability as well as seeing versus transparency driven changes. In addition, the coherent time-series structure in the data in Figure~\ref{fig: combined_all_fibers} suggests that there truly is a time-variable component to the science observations. We have further explored the origin of the $\vec b$ signal through comparisons with independent-but-simultaneous measurements from a guider camera, with results confirming that our model is able to capture real atmospheric variations. Thus, the time evolution we measure is not merely the result of poorly-calibrated or misunderstood data, and is instead physically motivated.

\section{Conclusion} 
\label{sec:Conclusion}

This work proposes a new statistical model for extracting optimal time-varying fluxes from NIR detectors (Section~\ref{sec:statistics}). This approach builds upon and generalizes previous models that were developed for space-based observations where sources are typically constant in flux during an exposure. Now, ground-based NIR observations that experience changing atmospheric conditions can measure unbiased fluxes with optimal uncertainties.

We validate our model by comparing to a variety of realistic synthetic datasets, showing that we can recover the input truth across a range of flux levels (Section~\ref{sec:fit_repeat_obs}) and amplitudes of flux variability (Section~\ref{sec:fit_repeat_obs_b_amp}). We find that APOGEE-like observations should use the time-dependent model over the time-independent model when the amplitude of flux variation is $\geq3.5$\% to avoid flux-dependent biases with magnitudes up to 120\% of the flux uncertainty (Figures~\ref{fig:synthetic_repeat_model_bias} to \ref{fig:synthetic_repeat_model_bias_levels}). 

We next show that our new model performs well on real data from the northern APOGEE spectrograph (Section~\ref{sec:apogee_data}). The time-variable model is able to describe data both in cases where the flux appears to be changing (Figure~\ref{fig: combined_all_fibers}) as well as with time-stable calibration observations (Figures~\ref{fig: model with calibration data} and \ref{fig:calibration_b_vect}). We confirm that the bulk of the time variability agrees with our model's wavelength independence assumption (Figures~\ref{fig: chi with real data} and \ref{fig: b_vector_compare}). We also explore the origin of these flux changes through comparisons with independent guider camera measurements of the atmospheric conditions, finding that the time evolution we measure is physically motivated by a variable sky (Figures~\ref{fig:guiderComp_trans_change}-\ref{fig:guiderComp_no_change}). For real APOGEE data, accounting for time-variability significantly improves data-model comparisons and produces unbiased fluxes.

% Based on assumptions and models similar to \citet{Brandt_2024}, it introduces a b vector within \citet{Brandt_2024} model to capture cloud factor variations. Tests on synthetic data demonstrate that in all three high, moderate, and low flux cases, the CDF of the empirical $\chi^2$ distribution of the b vector aligns with the theoretical CDF, although some underestimate or overestimate existes. The fitted flux z-score follows a standard normal distribution for all three cases, indicating the robustness and unbiasedness of this method. 

% When applied to APOGEE calibration data, we validate the extracted b vector as highly similar to the expected b vector (vector one). In testing with real APOGEE data, the $\chi^2$ distribution showed significant improvement compared to the decade-old established method \citep{Rauscher_2007}, and we verified that the b vector shows some dependence across different wavelengths, but this is acceptable.

% It is noteworthy, however, that the b vector exhibits significant fluctuations in the actual data. This may stem from issues in the processing of APOGEE data or inherent problems with the APOGEE instrument itself, representing an important direction for future research. 

We plan to incorporate this model into a new APOGEE reduction pipeline that is under active development: \texttt{ApogeeReduction.jl}.\footnote{The GitHub repository for the pipeline can be found at \url{https://github.com/andrew-saydjari/ApogeeReduction.jl}} Ongoing and future work will focus on tackling a few outstanding questions before we can implement this approach at scale. This includes understanding the best way to measure $\vec b$ from a large collection of pixels (i.e. $2048\times 2048$~pixel$^{2}$ detectors) that are potentially seeing different atmospheric conditions, how best to measure time-variability in fibers with very faint pixels, and optimizing the code to prevent the introduction of new bottlenecks into the reduction process. 

\begin{acknowledgements}

\bl{The authors thank the anonymous referee for comments that helped improve the clarity of this paper. The authors also}  thank Adam Wheeler for helpful conversations that improved the body of this work. 
    
K.A.M. acknowledges support from the University of Toronto’s Eric and Wendy Schmidt AI in Science Post-doctoral Fellowship, a program of Schmidt Sciences.

A.K.S. acknowledges support for this work was provided by NASA through the NASA Hubble Fellowship grant HST-HF2-51564.001-A awarded by the Space Telescope Science Institute, which is operated by the Association of Universities for Research in Astronomy, Inc., for NASA, under contract NAS5-26555.

J.G.F-T gratefully acknowledges the grants support provided by ANID Fondecyt Postdoc No. 3230001 (Sponsoring researcher), the Joint Committee ESO-Government of Chile under the agreement 2023 ORP 062/2023, and the support of the Doctoral Program in Artificial Intelligence, DISC-UCN.

Funding for the Sloan Digital Sky Survey V has been provided by the Alfred P. Sloan Foundation, the Heising-Simons Foundation, the National Science Foundation, and the Participating Institutions. SDSS acknowledges support and resources from the Center for High-Performance Computing at the University of Utah. SDSS telescopes are located at Apache Point Observatory, funded by the Astrophysical Research Consortium and operated by New Mexico State University, and at Las Campanas Observatory, operated by the Carnegie Institution for Science. The SDSS web site is \url{www.sdss.org}.

SDSS is managed by the Astrophysical Research Consortium for the Participating Institutions of the SDSS Collaboration, including the Carnegie Institution for Science, Chilean National Time Allocation Committee (CNTAC) ratified researchers, Caltech, the Gotham Participation Group, Harvard University, Heidelberg University, The Flatiron Institute, The Johns Hopkins University, L'Ecole polytechnique f\'{e}d\'{e}rale de Lausanne (EPFL), Leibniz-Institut f\"{u}r Astrophysik Potsdam (AIP), Max-Planck-Institut f\"{u}r Astronomie (MPIA Heidelberg), Max-Planck-Institut f\"{u}r Extraterrestrische Physik (MPE), Nanjing University, National Astronomical Observatories of China (NAOC), New Mexico State University, The Ohio State University, Pennsylvania State University, Smithsonian Astrophysical Observatory, Space Telescope Science Institute (STScI), the Stellar Astrophysics Participation Group, Universidad Nacional Aut\'{o}noma de M\'{e}xico, University of Arizona, University of Colorado Boulder, University of Illinois at Urbana-Champaign, University of Toronto, University of Utah, University of Virginia, Yale University, and Yunnan University.
    
\end{acknowledgements}

\vspace{5mm}
\facilities{Sloan (APOGEE-N), Du Pont (APOGEE-S)}

\software{\texttt{ApogeeReduction.jl} (in prep.), \texttt{fitramp} \citep{Brandt_2024}, \texttt{jupyter} \citep{jupyter_citation}, \texttt{matplotlib}  \citep{matplotlib_citation}, \texttt{numpy} \citep{numpy_citation}, \texttt{pandas} \citep{pandas_citation_2010,pandas_citation_2020}, \texttt{scipy} \citep{scipy_citation}}

\bibliography{main}{}
\bibliographystyle{aasjournal}

\appendix
% \onecolumngrid
\twocolumngrid

\section{Gibbs versus Linearized Fitting} \label{sec:method_comp_appendix}

\begin{figure*}[b]
    \centering
    \includegraphics[width=\linewidth]{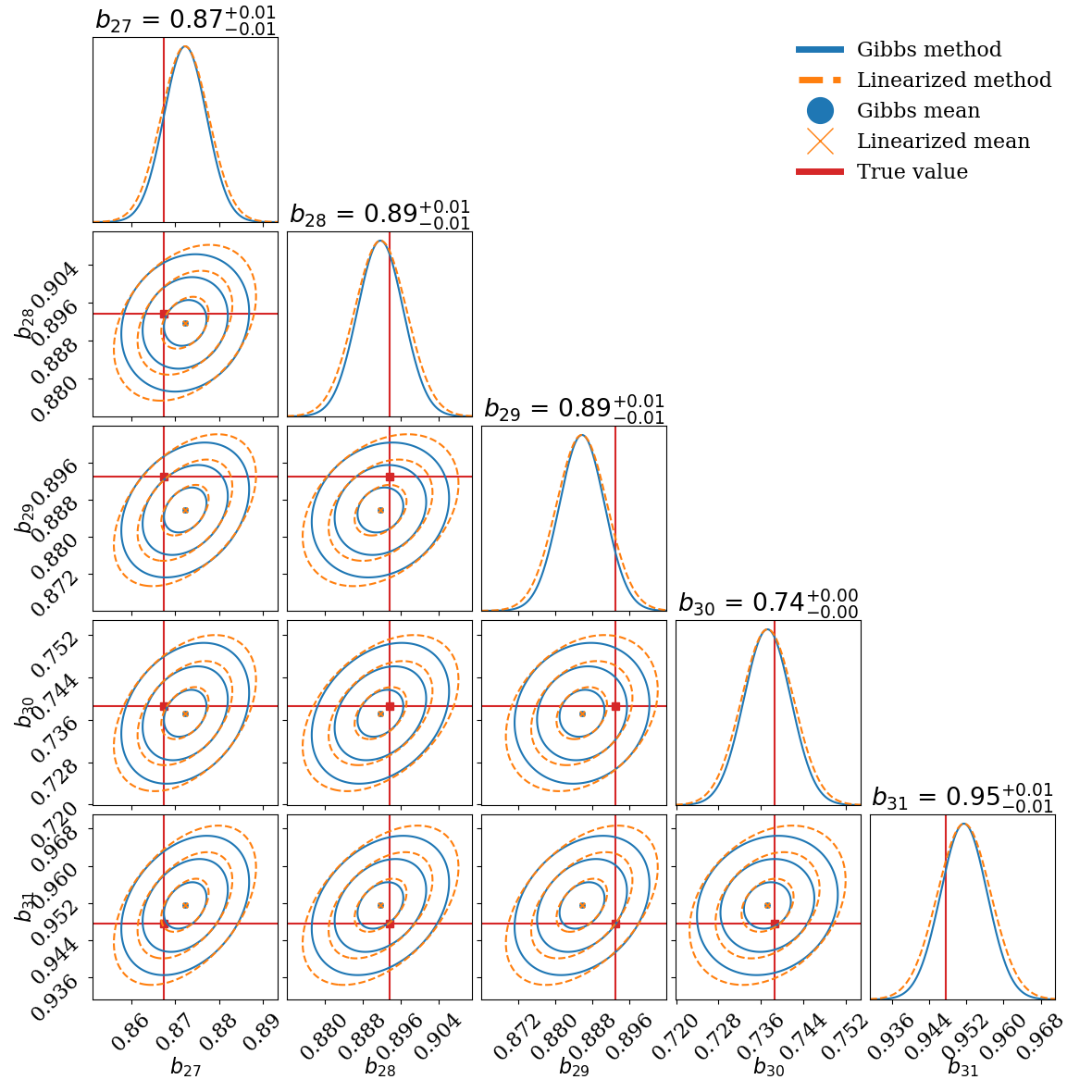}
    \caption{Corner plot comparing the multidimensional relationships between the last five $\vec b$ elements obtained from the Gibbs method (blue lines) and the linearized model (orange lines). The ellipses show the 68\%, 95\% and 99\% contours defined by the posterior distribution from each method, while the red lines show the input truth used to generate the synthetic data. The titles above the diagonals display the mean and 68\% credible region as defined by the Gibbs sampling method. Both methods agree quite well with the truth and return almost identical mean estimates. The key difference between the approaches comes from the size and shapes of the posterior distributions, though they still broadly agree with each other.}
    \label{Fig:gibbs_vs_linear}
\end{figure*}

Here we compare the results of fitting synthetic data with the Gibbs method using the posterior conditionals in Equation~\ref{eq:posterior_conditionals} versus the linearized model defined by Equation~\ref{eq:linear_posterior}. The precise data we fit are from the high flux case described in Section~\ref{sec:fit_single_obs} and Figure~\ref{fig:synthetic_single_output_comp}.

\begin{figure*}[b]
    \centering    \includegraphics[trim={1cm 1cm 2cm 2cm},width=\linewidth]{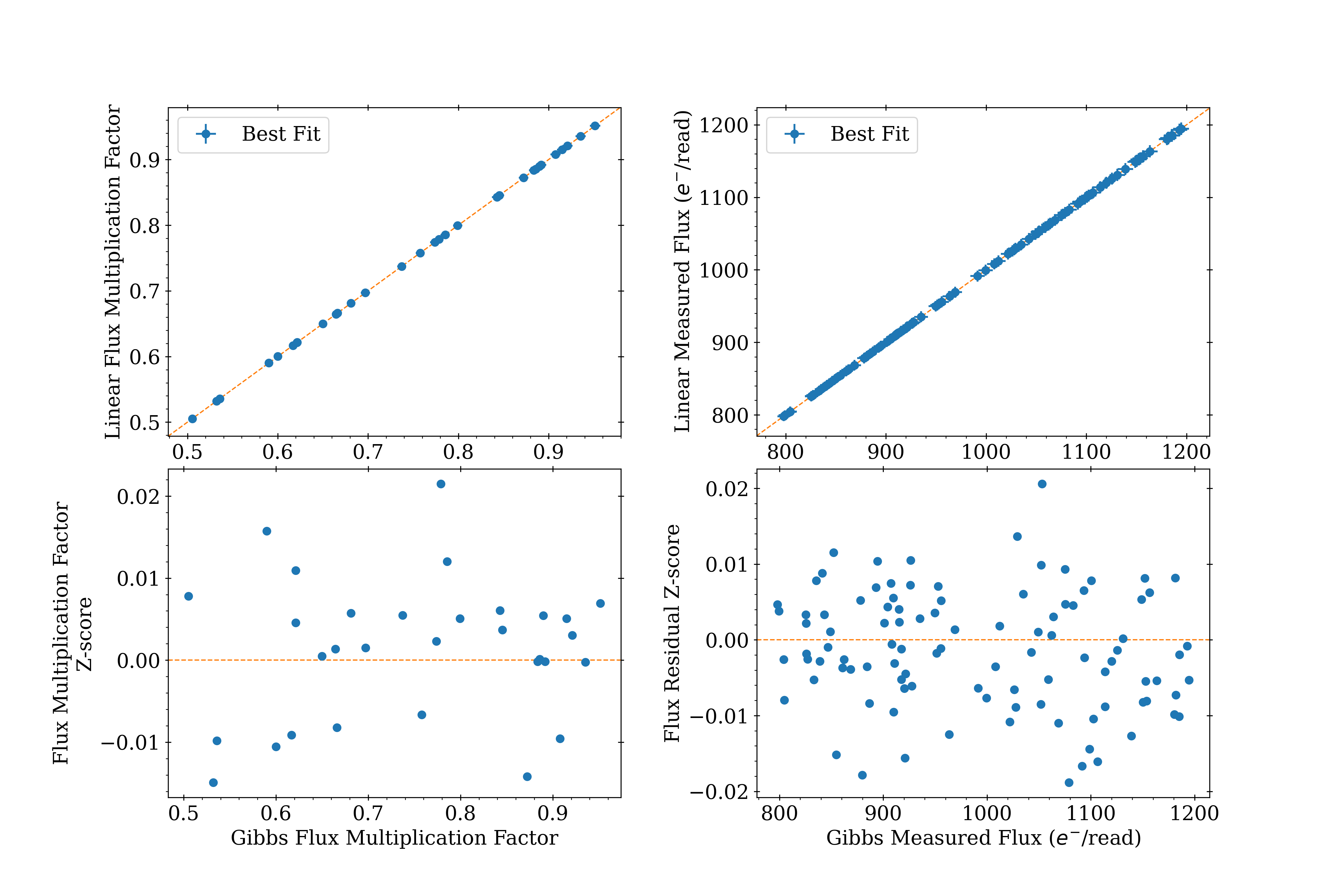}
    \caption{Comparison of the best-fit mean $\vec b$ and $\vec f$ using both the linearized model and Gibbs method. Like Figure~\ref{fig:synthetic_single_output_comp}, the left panels focus on $\vec b$ while the right focus on $\vec f$. The top panels show the direct one-to-one comparison of the measurements of the different techniques, and the bottom shows the residuals scaled by the quadrature-combined-uncertainty. We see that the z-scores are all very close to 0 \bl{(typically smaller than 1\% of a corresponding uncertainty unit)}, showing that the different techniques provide mean estimates that are almost identical to one another.}
    \label{Fig:gibbs_vs_linear_mean_comparison}
\end{figure*}

For the linearized model, the posterior distribution is defined directly by Equation~\ref{eq:linear_posterior}. For the Gibbs method, we estimate the full posterior $(\vec f, \vec b \mid \mathrm{data})$ by drawing samples from the posterior conditionals $(\vec f \mid \vec b, \mathrm{data})$ and $(\vec b \mid \vec f, \mathrm{data})$ and then taking a mean and covariance of the $(\vec f, \vec b \mid \mathrm{data})$ samples to define a multivariate Gaussian that can be easily compared to the linearized results. Figure~\ref{Fig:gibbs_vs_linear} compares the two approaches when focusing on the posterior distributions of the last five elements of the $\vec b$ vector. This corner plot shows that the mean estimates are virtually identical, but that there is some difference in the width and correlations of the multivariate posteriors. In this case, the Gibbs method (blue) tends to produce narrower distributions than the linearized model (orange). However, as discussed in Section~\ref{sec:statistics}, we are typically only concerned with the mean or MAP $\vec b$ when trying to extract unbiased fluxes; the uncertainty and correlations of the $\vec b$ are not propagated to our best fit $(\vec f \mid \vec b)$ measurements, as discussed in Section~\ref{sec:caveats}. The mean $\vec b$ for all read differences and $\vec f$ for all synthetic pixels in this analysis are directly compared in Figure~\ref{Fig:gibbs_vs_linear_mean_comparison}. The top panels show that both methods recover almost identical measurements, and the bottom panels show that these mean estimates are extremely similar when compared to the size of their uncertainties. The fact that both methods agree so closely in their $\vec b$ mean estimates -- that is, typically within 1\% of their uncertainty -- shows that neglecting the robust-but-slow Gibbs method in favor of the faster linearized model is likely to have a negligible impact on final results. \bl{When we repeat the analysis but use $\vec b = \vec 1$ and compare the extracted fluxes to the time-independent fits, we find that the flux differences between the Gibbs and linearized method are approximately five times smaller than the difference between either method and the time-independent result. This is further evidence that the choice of Gibbs versus linear is not a substantial source of error.}

% \begin{figure}[ht]
%     \centering
%     \begin{minipage}{0.48\linewidth}
%         \centering
%         \includegraphics[width=\linewidth]{figures/appendix 1.png}
%     \end{minipage}
%     \hfill
%     \begin{minipage}{0.48\linewidth}
%         \centering
%         \includegraphics[width=\linewidth]{figures/appendix 2.png}
%     \end{minipage}
%       \caption{GIBBS VS. LINEARIZED FITTING}
%     \label{Fig:Appendix}
% \end{figure}

\end{document}